\documentclass[sigconf]{acmart}
\usepackage{graphicx} % Required for inserting images
\usepackage{tikz}
\usepackage{multirow}
\usepackage{algorithm}
\usepackage{algpseudocode}
\usepackage{enumitem}

\setcopyright{none}
\renewcommand\footnotetextcopyrightpermission[1]{}
\acmConference[SPAA '27]{ACM Symposium on Parallelism in Algorithms and
  Architectures}{June 5--11, 2027}{Atlanta, GA, USA}

\newcommand{\eps}{\varepsilon}
\newcommand{\kh}{\widehat k}
\newcommand{\ellh}{\widehat{\ell}}
\newcommand{\rh}{\widehat r}
\newcommand{\ceil}[1]{\left\lceil #1\right\rceil}
\newcommand{\floor}[1]{\left\lfloor #1\right\rfloor}
\newcommand{\bfo}{DBP} % dynamic balanced pressures (formerly BFO)

\newcommand{\myparagraph}[1]{\smallskip\noindent {\bf #1.}}

\begin{document}

\title{Batch-Dynamic Coreness and Applications with Logarithmic Work via Balanced Pressures}
\author{Quinten De Man}
\email{deman@umd.edu}
\affiliation{%
  \institution{University of Maryland}
  \city{College Park}
  \state{Maryland}
  \country{USA}}
\author{Laxman Dhulipala}
\email{laxman@umd.edu}
\affiliation{%
  \institution{University of Maryland}
  \city{College Park}
  \state{Maryland}
  \country{USA}}
\author{Kishen N Gowda}
\email{kishen19@umd.edu}
\affiliation{%
  \institution{University of Maryland}
  \city{College Park}
  \state{Maryland}
  \country{USA}}
\author{Zhongqi Wang}
\email{zqwang@umd.edu}
\affiliation{%
  \institution{University of Maryland}
  \city{College Park}
  \state{Maryland}
  \country{USA}}

\begin{abstract}
We give a parallel algorithm that maintains a one-sided
$(2+\eps)$-approximation of all core numbers under batches of edge
insertions and deletions. For fixed $\eps>0$, the expected amortized work
is $O(\log n)$ per edge update, and each batch has
deterministic polylogarithmic depth. This improves on the previously known $(4+\eps)$-approximation and $O(\log^2 n)$ amortized work bounds. This approximation factor
approaches the factor-two limit for polylogarithmic update and query time
under the OMv conjecture.

Our algorithm is based on assigning each vertex a ``pressure'' and splits each edge's unit
load smoothly between its endpoints according to their pressures.
After each batch of updates, our dynamic algorithm adjusts the pressures of vertices in parallel to restore balance between the pressures and loads.
We show that by maintaining balanced pressures we can obtain good approximations of core
numbers. 
Our construction also gives a $(1+\eps)$-approximation of the densest-subgraph
value with logarithmic expected amortized work and supports batch-dynamic maximal matching,
clique counting, forest decomposition, and coloring in low-arboricity graphs with improved or matching work bounds.

We give a practical implementation of our batch-dynamic coreness algorithm and find that it is able to achieve smaller empirical error than the state-of-the-art parallel batch-dynamic implementation, while running faster. 
On six real-world graphs, it runs $32\times$ faster than
that implementation in geometric mean and $4.1\times$ faster than a state-of-the-art heuristic baseline. 
On three graphs with up to $1.8$ billion edges, it is $9.2\times$ faster than the heuristic baseline.
\end{abstract}
    
\maketitle
% preprint: no running header naming the venue, page numbers only
\pagestyle{plain}

\section{Introduction}\label{sec:intro}

The core number of a vertex measures how well it is connected within a
cohesive part of a graph. Formally, the \emph{$k$-core} is the largest subgraph
in which every vertex has degree at least $k$, and the \emph{core number} (or \emph{coreness})
$k(v)$ is the largest $k$ for which $v$ belongs to the $k$-core. These nested
subgraphs give a simple description of network structure~\cite{Seidman83}
and help identify influential spreaders~\cite{Kitsak10}, among many other practical applications~\cite{AHDBV05,SG10,BH03,RV15}.

The exact core numbers of every vertex can
be computed in $O(n+m)$ time by repeatedly peeling a vertex of minimum
degree~\cite{MatulaBeck83,BZ03}.
However, maintaining coreness dynamically as the graph changes is much harder, as adding one edge to close a path into a cycle changes every core
number from $1$ to $2$. 
Thus, an algorithm that explicitly maintains exact core
numbers must sometimes update \emph{every} vertex in response to one
edge. 
Allowing for approximation in the core numbers allows algorithms to defer changes until they become
significant, making it possible to achieve much lower work in theory and in practice.

In this paper, we study coreness in the \emph{fully dynamic} setting, with both edge insertions and
deletions, and allow updates to arrive in batches of any size. 
In particular, we are interested in the \emph{parallel batch-dynamic setting}, where the goal is to process a batch with low \emph{work} per edge update
and low \emph{depth} (i.e., the length of the longest chain of dependencies). 
This setting is a strict generalization of the classical dynamic setting, where we have a batch size of one.

\myparagraph{Previous work}
Sun et al.~\cite{SCS20} gave the first algorithm for approximate core numbers with polylogarithmic update time. They provide a $(4+\eps)$-approximation with $O(\log^2 n)$ amortized update time for constant $\eps>0$.
Liu et al.~\cite{LSYDS22} introduced the parallel level data structure
(PLDS), obtaining the same approximation guarantee with
$O(\log^2 n)$ amortized work per update and polylogarithmic depth per batch.
More recently, Ghaffari and Koo~\cite{GK25} obtained polylogarithmic
\emph{worst-case} work per update, again with a $(4+\eps)$-approximation
guarantee. Table~\ref{tab:comparison} compares the bounds. We state all
approximation factors in the one-sided convention
$k(v)/c\le\kh(v)\le k(v)$, where $k(v)$ is the true core number and
$\kh(v)$ is its estimate; Section~\ref{sec:prelims} explains the conversion
between conventions used in this paper and prior papers.

\begin{table*}[t]
\caption{Fully dynamic core-number bounds for constant $\eps>0$. A factor
$c$ means $k(v)/c\le\kh(v)\le k(v)$ after rescaling estimates and
reparameterizing accuracy. Work is per edge update; depth is per batch.
The work and depth bounds of PLDS and Ghaffari--Koo hold w.h.p.; our work
bound is in expectation and our depth bound is deterministic. The two
Ghaffari--Koo rows describe update types of the same fully dynamic algorithm.}
\label{tab:comparison}
\centering
\small
\begin{tabular}{@{}lllll@{}}
\toprule
Algorithm & One-sided factor & Work per update & Work guarantee & Depth per batch \\
\midrule
Sun--Chan--Sozio~\cite{SCS20} & $4+\eps$ & $O(\log^2 n)$ & Amortized & Sequential \\
Liu et al. (PLDS)~\cite{LSYDS22} & $4+\eps$ & $O(\log^2 n)$ & Amortized & $O(\log^2 n\log\log n)$ \\
Ghaffari--Koo~\cite{GK25}, insertions & $4+\eps$ & $O(\log^9 n)$ & Worst-case & $O(\log^7 n)$ \\
Ghaffari--Koo~\cite{GK25}, deletions & $4+\eps$ & $O(\log^8 n)$ & Worst-case & $O(\log^6 n)$ \\
This paper & $2+\eps$ & $O(\log n)$ & Expected amortized & $O(\log^4 n)$ \\
\bottomrule
\end{tabular}
\end{table*}

\myparagraph{Our result}
We give a simple and practical batch-dynamic algorithm that improves both the approximation factor and
the amortized work. For every $0<\eps\le1$ with $1/\eps\le n^{O(1)}$, it
maintains estimates satisfying
\[
 \frac{k(v)}{2+\eps}\ \le\ \kh(v)\ \le\ k(v)
 \qquad\text{for every vertex }v
\]
after every batch, using $O(\eps^{-5}\log n)$ expected amortized work per
update\footnote{The expectation comes only from parallel semisorting and
hash-table operations. The pressure updates and potential analysis are
deterministic. A deterministic sequential implementation takes
$O(\eps^{-5}\log n)$ amortized time per update
(Appendix~\ref{app:sequential}). We focus on the parallel setting.}
and deterministic $O(\eps^{-3}\log^4 n)$ depth per batch.
For constant $\eps$, the space is $O(m+n)$.
Theorem~\ref{thm:main} gives the full statement. Thus the algorithm reduces
the work from two logarithms to one while improving the previously stated
approximation factor from about four to about two. Equivalently, rescaling
our estimates gives a two-sided $\sqrt{2+\eps}$ approximation.

The factor of two is also a conditional limit on what a fast dynamic
algorithm can achieve. Couto and Fernandes~\cite{CF25} show that, under the
Online Matrix--Vector Multiplication (OMv) conjecture, no algorithm with
polynomial preprocessing and polylogarithmic update and query time can
achieve a one-sided $(2-\delta)$ approximation for any fixed $\delta>0$.
Our approximation approaches this threshold while using logarithmic
expected amortized work.

\myparagraph{Densest subgraph and other applications}
Our data structure also gives a fully dynamic $(1+\eps)$-approximation of
the densest-subgraph value with $O(\log n)$ amortized update
time and constant query time, for fixed $\eps>0$. To our knowledge, this is
the first logarithmic update bound for this approximation guarantee,
improving the $O(\log^2 n)$ amortized bound of Chekuri and
Quanrud~\cite{CQ22}. Their logarithmic-time variant allows
an additive error; pure multiplicative approximations with polylogarithmic
worst-case updates are also known~\cite{CCHHQRSch24,GK25}.
The batch-dynamic algorithm achieves $O(\log n)$
expected amortized work per update and polylogarithmic depth per batch.
It also reports the vertices of a $(1+\eps)$-approximate
densest subgraph $G[S]$ in $O(|S|+\log^2 n)$ work and $O(\log n)$ depth,
with the same update bounds.
Appendix~\ref{app:density} gives the proof.

The data structure also improves or matches work bounds in the
low-arboricity framework for orientations, maximal matching, clique
counting, forest decompositions, and coloring.
Appendix~\ref{app:applications} gives the reductions and comparisons.

\myparagraph{Experimental results}
We implement the algorithm and compare it with PLDS and PLDSOpt on the six
social and web graphs and the three largest graphs of~\cite{LSYDS22}, using
insertion, deletion, and mixed batches of $10^6$ updates on 60 threads
(Section~\ref{sec:exp}). The implementation uses smaller parameters and a
simpler rule for restoring balance than Theorem~\ref{thm:main}. Its
accuracy remains provable for every parameter setting, while its work is
measured. With parameters that give a guaranteed two-sided error bound
of at most $3.8$, below the $4.2$ guaranteed for PLDS, it is faster than both PLDS
and PLDSOpt on every input. On the six social and web graphs, it is
$32\times$ faster than PLDS and $4.1\times$ faster than PLDSOpt in
geometric mean (from $1.9\times$ to $10.3\times$), and its average error is
lower than that of PLDSOpt on 14 of the 18 combinations of graph and update
type. On three graphs with up to $1.8$ billion edges, it is $9.2\times$
faster than PLDSOpt.
Recomputing all core numbers from scratch after each batch is faster only
when a batch changes a large part of the graph, about a fifth of the edges
of \emph{livejournal}.

\subsection{Technical Overview}

We now give an overview of the main ideas.
Each edge in the undirected graph contributes one unit of load that is divided (fractionally) between its endpoints.
The portion assigned to an endpoint is its \emph{share} of the edge.
Together, these shares form a \emph{fractional orientation}.
The \emph{load} $\ell(v)$ at vertex $v$ is the sum of the shares it receives
from its incident edges.
To determine the edge shares,
we store a positive number $p(v)$, called the \emph{pressure}, at each
vertex $v$. For an edge $uv$ with $p(u)\le p(v)$, vertex $v$ receives
a share of $\frac12(p(u)/p(v))^s$ and $u$ receives the remaining share
$1-\frac12(p(u)/p(v))^s$. We set $s$ later to obtain the approximation
guarantee.

\myparagraph{Algorithm}
We call the pressures \emph{balanced} if the load $\ell(v)$ lies between
$(1-\eps/8)p(v)$ and $(1+\eps/8)p(v)$ at every non-isolated vertex.
Here $\eps$ is the same accuracy parameter as in the $(2+\eps)$-approximation guarantee.
Our algorithm maintains balanced pressures after every batch
and uses them to estimate core numbers and support the other applications.
We first delete the edges in the batch, reducing the loads at their
endpoints. These deletions can leave pressures too high relative to loads,
so the parallel rounds described below lower pressures as needed to restore
balance.
Before inserting new edges, we raise each affected vertex's pressure to a
constant fraction of its number of new incident edges, unless its pressure
is already that high.
We then insert the new edges, reset isolated vertices to the minimum
pressure, and restore balance over parallel rounds.
In each round, a vertex whose load is sufficiently above its pressure multiplies its pressure by a factor $\beta>1$ to reduce its load. Similarly, a vertex whose load is sufficiently below its pressure divides its pressure by $\beta$ to increase its load.
All of the updates within a round use the loads at the start of the round.
We choose $\beta=1+\Theta(\eps^3/\log n)$, so each share changes by only an
$O(\eps)$ fraction in a round. Section~\ref{sec:batch} explains the choice of step size.
We repeat these rounds until balance is restored.
We prove that balanced pressures approximate core
numbers and bound the work needed to restore balance.

\myparagraph{Balanced pressures approximate core numbers}
We show that, at every non-isolated vertex $v$, a balanced pressure $p(v)$ lies between $k(v)/2$ and $k(v)$ to within a multiplicative $(1\pm O(\epsilon))$ factor.

For the lower bound on $p(v)$, let $k=k(v)$ and choose a vertex $u$ of
minimum pressure in the $k$-core.
Each of its at least $k$ neighbors in the core has pressure at least $p(u)$, so $u$ receives at least half the load of each edge to these neighbors.
Its load is therefore at least $k/2$, and balance gives $p(u)\ge(1-O(\eps))k/2$.
Since $v$ belongs to this core, $p(v)\ge p(u)$, giving the lower bound
for $p(v)$.

For the upper bound on $p(v)$, consider the non-isolated vertices in the prefix ending at $v$ in a peeling order.
Each has at most $k(v)$ neighbors appearing after it in the peeling order.
We weight each vertex by a large power of its pressure and consider the sum of their weighted loads. We show that this sum is at most approximately $k(v)$ times the sum of the weights. 
Specifically, the share rule bounds each edge's contributes to this weighted sum by approximately the weight of its lower-pressure endpoint, which is no greater than the weight of the endpoint peeled first.
Thus we can charge each edge to its earlier endpoint, giving the claimed bound on the weighted loads.
Using balance, this becomes a bound on the weighted average pressure. Our choice of exponent makes this average close to the largest pressure in the prefix, which is at least $p(v)$, and thus yields the bound $p(v)\le(1+O(\eps))k(v)$.
We obtain the estimated core number $\kh(v)$ by dividing $p(v)$ by a
suitable factor $1+O(\eps)$, so that $\kh(v)\le k(v)$.

\myparagraph{Amortized work}
Our updates follow the multiplicative-weights approach~\cite{AHK12,BG19},
adjusting pressures by small multiplicative factors in response to
imbalances. We bound their cost using a nonnegative potential that accounts
for both edge updates and the work of restoring balance.
For each graph $G$, we construct a function $F_G$ of the pressures that is convex in their logarithms.
Our potential is the current value of $F_G$ minus its minimum over \emph{any} pressure assignment, keeping $G$ fixed.
We show that each moving vertex changes its pressure in a direction that decreases $F_G$.
We choose steps small enough that, after all updates in a round, a vertex that raised its pressure still has load above pressure, and one that lowered its pressure still has load below pressure.
These inequalities and convexity give a lower bound on the potential decrease over the whole round.

We next bound how much edge updates can increase the potential.
We show that the pressure assignment minimizing $F_G$ is ``perfectly balanced'', i.e., each non-isolated vertex's load equals its pressure.
Since each edge contributes one unit of load, the pressures of the non-isolated vertices sum to the number of edges.
We also show that inserting an edge never decreases the minimizing pressures and increases their sum by at most one.
Using this bound and the corresponding bound for deletions, we show that each batch increases the potential by at most $O(1)$ per edge update before the parallel rounds.

The potential starts at zero on the empty graph, and the batches add $O(T)$ in total over a sequence of $T$ edge updates.
Since the potential remains nonnegative, its total decrease over all rounds is at most $O(T)$.
For fixed $\eps$, we show that each round's expected work is at most $O(\log n)$ times its potential decrease, giving $O(T\log n)$ expected work over all rounds.
%
%Applying the edge updates and the pressure increases before insertion fits within the same bound.

\myparagraph{Parallel rounds and aggregation}
To bound the depth per batch, we analyze both the number of rounds and the depth of each round.
For the round bound, we use a tighter balance tolerance to select moving vertices than to decide when to stop.
With sufficiently small steps, a vertex still unbalanced after $t$ rounds must have increased or decreased its pressure by a factor of $\beta^t$.
Since pressures remain between a positive constant and $O(n)$, our choice of $\beta$ limits a batch to $O(\log^2 n)$ rounds for fixed $\eps$.

To implement a round efficiently, we update the loads without scanning every neighbor of a vertex that is updating its pressure (a ``moving'' vertex).
When a vertex changes its pressure, all shares it receives from much lower-pressure neighbors scale by the same factor.
We store their total contribution in counters and update it with simple arithmetic, without scanning those neighbors.
By allowing a small approximation error, we can also avoid updating those neighbors' stored loads while the pressure gap remains large.
For fixed $\eps$, the moving vertex receives at least a constant share of each edge we scan.
Its load is the sum of its edge shares, so the number of scanned edges is at most a constant times its load.
The counter updates and neighbor processing have deterministic polylogarithmic depth.
Section~\ref{sec:alg} establishes the properties of balanced pressures
and the accuracy of the estimates. Section~\ref{sec:ds} gives the update
algorithm, its data structure, and the work and depth analysis.

\section{Preliminaries}\label{sec:prelims}

\myparagraph{Graphs}
We consider undirected graphs without self-loops or parallel edges. For a graph
$G=(V,E)$ we write $n=|V|$ and $m=|E|$. The vertex set is fixed and the edge set
changes over time, so $m$ always denotes the current number of edges. For a
vertex $v$, $N(v)$ is its set of neighbors and $\deg(v)=|N(v)|$ is its degree.
We write $\ln$ for the natural logarithm and $\log$ for the
base~$2$ logarithm.

\myparagraph{Cores and core numbers}
For an integer $k\ge0$, the \emph{$k$-core} of $G$ is the largest subgraph of
$G$ in which every vertex has degree at least $k$~\cite{Seidman83}. It is well
defined because the union of two subgraphs of minimum degree at least $k$ again
has minimum degree at least $k$. The \emph{core number} $k(v)$ of a vertex $v$,
also called its \emph{coreness}, is the largest $k$ such that $v$ belongs to the
$k$-core. Equivalently, $k(v)$ is the largest $k$ such that $v$ lies in some
subgraph of $G$ of minimum degree at least $k$. A vertex without edges has core
number~$0$.

All core numbers can be computed in $O(n+m)$ time by
\emph{peeling}~\cite{MatulaBeck83,BZ03}, i.e., by repeatedly deleting a vertex of minimum
degree from the remaining graph. Our analysis uses the following folklore
property of the order in which peeling deletes the vertices.

\begin{lemma}[Peeling order]\label{lem:peel}
Repeatedly delete a vertex of minimum degree from the remaining graph,
breaking ties by choosing the smallest vertex ID, until no vertex is left.
Let $v_1,\dots,v_n$ be the vertices in the order of deletion. Then
$k(v_1)\le\dots\le k(v_n)$, and every $v_i$ has at most $k(v_i)$ neighbors among
$v_{i+1},\dots,v_n$.
\end{lemma}

\myparagraph{Approximate core numbers}
An algorithm for approximate core numbers produces an \emph{estimate} $\kh(v)$ for
every vertex~$v$. Following Liu et al.~\cite{LSYDS22}, the estimates form a
\emph{$c$-approximation}, for a number $c\ge1$, if
\[
 \frac{k(v)}{c}\ \le\ \kh(v)\ \le\ c\,k(v)\qquad\text{for every vertex }v.
\]
The estimates of our algorithm are never larger than the core numbers. We call
estimates a \emph{one-sided $c$-approximation} if
\[
 \frac{k(v)}{c}\ \le\ \kh(v)\ \le\ k(v)\qquad\text{for every vertex }v.
\]
We can easily convert one form into the other. Multiplying a one-sided
$c$-approximation by $\sqrt c$ gives a $\sqrt c$-approximation, and dividing a
$c$-approximation by $c$ gives a one-sided $c^2$-approximation. Both conversions
keep the ratio between the largest and the smallest value that the guarantee
allows for $\kh(v)$; this ratio is $c$ for a one-sided $c$-approximation and
$c^2$ for a $c$-approximation.

\myparagraph{Batch-dynamic updates}
Initially the graph is empty, and is updated over a sequence of \emph{batch updates}. 
Formally, a batch is a pair $(D,I)$, where $D\subseteq E$ is a set of edges to delete and
$I$ is a set of edges not in $E$ to insert; it changes the edge set from $E$ to
$(E\setminus D)\cup I$. A batch given as an arbitrary list of insertions and
deletions can be brought into this form by grouping the operations by edge with
a semisort (defined below) and keeping, for every edge, the outcome of its last
operation. Each inserted or deleted edge is an \emph{update}. After every batch,
the algorithm holds an estimate for every vertex.

Work bounds are per update and amortized over the sequence: an algorithm takes
$O(w)$ amortized work per update if every sequence of batches with $T$ updates
in total takes $O(T\,w)$ work. Depth bounds are per batch. A batch with a single
update is an ordinary dynamic update, so the bounds include the sequential
setting in which edges are inserted and deleted one at a time. A graph that is
given at the start is inserted as one batch.

\myparagraph{Model of computation}
We analyze our algorithms in the work-depth model, specifically the
binary-forking model with test-and-set of Blelloch et al.~\cite{BFGS20}, with
words of $O(\log n)$ bits. The \emph{work} of a computation is the total number of
instructions executed, and its \emph{depth} is the length of the longest chain
of dependent instructions. The model is equivalent to the PRAM in work, and
depth in it equals PRAM time up to a logarithmic factor~\cite{BFGS20}. A bound
holds \emph{with high probability} (w.h.p.) if it holds with probability at least
$1-n^{-c}$ for every constant $c\ge1$.

\myparagraph{Parallel primitives}
Given a sequence $A=(a_1,\ldots,a_\ell)$, \emph{prefix sums} returns
$(s_1,\ldots,s_\ell)$, where $s_i=\sum_{j=1}^{i}a_j$, and \emph{reduce}
returns $s_\ell$. Given a constant-time predicate $p$, \emph{filter} returns
the subsequence of $A$ satisfying $p$, preserving its order. Each primitive
takes $O(\ell)$ work and $O(\log\ell)$ depth~\cite{BGS10}.
\emph{Semisort} permutes $\ell$ keyed items into contiguous groups of equal
keys, with arbitrary order between groups. Assuming
constant-time random hashing, we use a version with $O(\ell)$ expected
work, $O(\ell)$ space, and deterministic $O(\log^2(\ell+2))$ depth.
A \emph{parallel dictionary} stores key-value pairs
and supports batched insertions, deletions, and lookups.
Our dictionaries use linear space and process $\ell$
operations in $O(\ell)$ expected amortized work, with deterministic
$O(\log n)$ depth for lookups and $O(\log^2 n)$ depth for updates.
These depth bounds use the fact that dictionary and batch sizes are
polynomial in $n$. Appendix~\ref{app:primitives} gives the implementations
and proves their expected work and deterministic depth and space bounds.

\section{Balanced Pressures and Core-Number Estimates}\label{sec:alg}

Assign each vertex $v$ a positive number $p(v)$, called its \emph{pressure}.
Each edge divides one unit of load into two \emph{shares}, one for each
endpoint, according to the ratio of their pressures. Endpoints with equal
pressures receive a share of $1/2$ each. When the pressures differ, the
endpoint with lower pressure receives the larger share. The \emph{load}
$\ell(v)$ of vertex $v$ is the sum of the shares it receives from its
incident edges.

We call pressures \emph{balanced} when every non-isolated vertex's load
is close to its pressure. We show that balanced pressures approximate
core numbers, so the update algorithm in Section~\ref{sec:ds} can maintain
accurate estimates by restoring balance after each batch. We first define
balance precisely.

\begin{definition}[Balanced pressures]\label{def:bfo}
Fix an exponent $s>0$ and a tolerance $0<\sigma<1/2$.
Given pressures $p(v)\ge1/4$ at all vertices, define the endpoint shares of
each edge $uv$ with $p(u)\le p(v)$ by
\begin{equation}\label{eq:pressure-shares}
 x_{uv}(v)=\frac12\left(\frac{p(u)}{p(v)}\right)^s,
 \qquad x_{uv}(u)=1-x_{uv}(v).
\end{equation}
The load at $v$ is $\ell(v)=\sum_{u\in N(v)}x_{uv}(v)$.
The pressures are \emph{balanced} if every non-isolated vertex satisfies
\begin{equation}\label{eq:balanced}
 (1-2\sigma)\,p(v)\ \le\ \ell(v)\ \le\ (1+2\sigma)\,p(v),
\end{equation}
and every isolated vertex has pressure $1/4$.
\end{definition}

The accuracy of balanced pressures for estimating the coreness values depends on $s$ and
$\sigma$. The following theorem chooses these parameters to obtain a
$(2+\eps)$-approximation and bounds the work, depth, and space used by the
data structure.

\begin{theorem}\label{thm:main}
Fix $n\ge2$ and a rational $0<\eps\le1$ whose numerator and denominator
are at most $n^{O(1)}$.
There is a parallel data structure for an initially edgeless graph on $n$
vertices that supports batches of edge insertions and deletions. After every
batch, it maintains core-number estimates $\kh(v)$ satisfying
\[
 \frac{k(v)}{2+\eps}\le\kh(v)\le k(v)
 \qquad\text{for every vertex }v.
\]
It maintains these estimates using balanced pressures
(Definition~\ref{def:bfo}) with $\sigma=\eps/16$ and
$s=\ceil{2\ln n/\sigma^2}$.

After initialization, processing any sequence containing $T$
edge updates takes $O(T\eps^{-5}\log n)$ expected work.
A batch uses $O(\eps^{-3}\log^2n)$ rounds and has depth
$O(\eps^{-3}\log^4n)$. These bounds are deterministic.
The space usage is $O(m+n/\eps+\eps^{-3}\log n)$ words.
\end{theorem}

Section~\ref{sec:defs} establishes the share properties and chooses the
exponent $s$. Section~\ref{sec:accuracy} proves the core-number guarantee
and derives the estimates. Section~\ref{sec:ds} gives an algorithm for
maintaining balanced pressures and proves its work, depth, and space bounds.

\subsection{Definitions}\label{sec:defs}

Throughout, let $n\ge2$ and $\sigma=\eps/16\le\frac1{16}$. We choose the
share exponent $s$ at the end of this subsection.

\myparagraph{Edge shares}
The shares defined in Equation~\eqref{eq:pressure-shares} have three
properties that we use throughout the analysis.
\begin{description}[labelindent=5pt]
\item[(S1)] The two shares of an edge sum to $1$, and the endpoint with the
smaller or equal pressure receives at least $\frac12$.
\item[(S2)] A vertex's share does not decrease when its neighbor's pressure grows
or its own pressure shrinks.
\item[(S3)] For any factor $a\ge1$, if only one endpoint
multiplies or divides its pressure by $a$, both shares change by a factor
of at most $a^s$. If both endpoints do so, both shares change by a factor
of at most $a^{2s}$.
\end{description}
(S1) and (S2) follow from the definition.
For (S3), fix one endpoint's pressure and multiply or
divide the other's by a factor $a\ge1$. If the pressures do not change
order, Equation~\eqref{eq:pressure-shares} shows that the smaller share
changes by a factor $a^s$. The other share changes by the same absolute
amount, but is at least as large before and after the change, so its
multiplicative change is no greater.

If the change reverses the pressure order, apply the argument first until
the pressures become equal, then from that point to the final pressure.
The two pressure factors multiply to $a$, so each share changes by a
factor of at most $a^s$ over the whole change. Applying this bound once
for each endpoint gives the factor $a^{2s}$.

\myparagraph{Loads and ratios}
By (S1), the loads $\ell(v)$ sum to $m$.
For a vertex with at least one edge, we call
$r(v)=\ell(v)/p(v)$ its \emph{ratio}.

\myparagraph{Choosing the exponent}
The exponent $s$ controls how much less load an endpoint receives when its
pressure is higher than its neighbor's. This matters for vertices whose degree
is much larger than their core number. For example, the center of an
$n$-vertex star has core number $1$, so our accuracy bound requires its
pressure to be at most $1+O(\eps)$. The balance constraint then requires its load to be
$O(1)$ as well. With $n-1$ incident edges, its average share must therefore
be $O(1/n)$.

To see how $s$ affects this load, suppose the leaves have pressure $1$ and
the center has pressure $1+\eps$. The center's load is
\[
 \frac{n-1}{2(1+\eps)^s}.
\]
At these pressures, keeping the load $O(1)$ requires
$(1+\eps)^s=\Omega(n)$, or $s=\Omega(\eps^{-1}\log n)$.
Our accuracy proof for general graphs uses the larger choice
$s=\Theta(\eps^{-2}\log n)$ below. In particular, for the remainder of the description of our analysis and algorithm, we set
\[
 s=\ceil{\frac{2\ln n}{\sigma^2}}.
\]

\subsection{Balanced Pressures Approximate Core Numbers}\label{sec:accuracy}

With the parameter choices above, balanced pressures approximate core
numbers. The following theorem upper and lower bounds $p(v)$ in terms of $k(v)$.

\begin{theorem}\label{thm:accuracy}
If the pressures are balanced, then every vertex $v$ with at least one edge
satisfies
\[
 \frac{k(v)}{2(1+2\sigma)}\ \le\ p(v)\ \le\ \frac{1+2\sigma}{1-2\sigma}\,k(v).
\]
\end{theorem}
\begin{proof}
Fix a non-isolated vertex $v$. For the lower bound, consider the
$k(v)$-core, which contains $v$ by the definition of the core number.
Choose a vertex $u$ with the smallest pressure in this core, so
$p(u)\le p(v)$. Each of the at least $k(v)$ neighbors of $u$ in the core
has pressure at least $p(u)$, so (S1) gives $\ell(u)\ge k(v)/2$. Balance
then gives
\[
 k(v)/2\le \ell(u)\le(1+2\sigma)p(u)\le(1+2\sigma)p(v),
\]
and dividing by $1+2\sigma$ gives the lower bound of the theorem.

For the upper bound, let $S$ consist of $v$ and all non-isolated
vertices that appear before $v$ in the peeling order. By
Lemma~\ref{lem:peel}, every $u\in S$ has $k(u)\le k(v)$ and at most $k(v)$
later neighbors. Charging each edge to its earlier endpoint shows that the
loads in $S$ sum to at most $k(v)|S|$, but this bounds only the average
load, which still allows individual pressures to be large. We therefore
multiply the load of each $u\in S$ by $p(u)^{\sigma s}$ before summing.
Because the exponent $\sigma s$ is large, this emphasizes the vertices
with the highest pressures, so the resulting bound applies to $p(v)$
itself rather than to the average. We first show that
\begin{align*}
 (1-2\sigma)\sum_{u\in S}p(u)^{\sigma s}p(u)
 &\le\sum_{u\in S}p(u)^{\sigma s}\ell(u)\\
 &\le\left(1+\frac\sigma2\right)k(v)\sum_{u\in S}p(u)^{\sigma s}.
\end{align*}
The first inequality follows by applying balance,
$\ell(u)\ge(1-2\sigma)p(u)$, to each term.

For the second inequality, we expand each load $\ell(u)$ in the
middle sum into the shares $u$ receives from its edges, and group the
resulting terms by edge. Consider an edge within $S$ with endpoint
pressures $a\le b$. Its two terms add up to
\[
 \begin{aligned}
 &a^{\sigma s}\bigl(1-\tfrac12(a/b)^s\bigr)+\tfrac12\,b^{\sigma s}(a/b)^s\\
 &\quad=a^{\sigma s}\Bigl(1+\tfrac12(a/b)^s\bigl((b/a)^{\sigma s}-1\bigr)\Bigr)\\
 &\quad\le\left(1+\frac\sigma2\right)a^{\sigma s}.
 \end{aligned}
\]
The inequality holds because the concavity of $x^\sigma$ gives
$(b/a)^{\sigma s}-1\le\sigma\bigl((b/a)^s-1\bigr)$, so the term added
to $1$ in the second line is at most
$\frac\sigma2\bigl(1-(a/b)^s\bigr)\le\frac\sigma2$. Since
$a^{\sigma s}\le b^{\sigma s}$, the edge's terms are at most
$(1+\sigma/2)\,p(u)^{\sigma s}$ for either endpoint $u$, and we charge
them to the endpoint that appears earlier in the peeling order. An edge
leaving $S$ has a single term, which is at most $p(u)^{\sigma s}$ for its
endpoint $u\in S$. That endpoint is the earlier one, because $S$ contains
all non-isolated vertices preceding $v$. Each $u\in S$ is charged for at
most $k(v)$ edges, one for each later neighbor, which proves the second
inequality.

To bound $p(v)$, write
\[
 A=\sum_{u\in S}p(u)^{\sigma s}p(u)=\sum_{u\in S}p(u)^{\sigma s+1}
\]
for the sum on the left, so the chain above gives
$(1-2\sigma)A\le(1+\sigma/2)k(v)\sum_{u\in S}p(u)^{\sigma s}$.
The largest pressures in $S$ dominate $A$, in particular,
$p(v)\le A^{1/(\sigma s+1)}$ because $v\in S$. So it suffices to upper bound
$A^{1/(\sigma s+1)}$.
The power mean inequality gives
\[
 \sum_{u\in S}p(u)^{\sigma s}
 \le |S|^{1/(\sigma s+1)}A^{\sigma s/(\sigma s+1)}.
\]
Substituting this bound for the sum of $p(u)^{\sigma s}$ gives
\[
 (1-2\sigma)A
 \le (1+\sigma/2)k(v)\,|S|^{1/(\sigma s+1)}A^{\sigma s/(\sigma s+1)}.
\]
Since $A>0$, dividing by $(1-2\sigma)A^{\sigma s/(\sigma s+1)}$ gives
\[
 A^{1/(\sigma s+1)}
 \le\frac{1+\sigma/2}{1-2\sigma}\,k(v)\,|S|^{1/(\sigma s+1)}.
\]
Using $v\in S$ and $|S|\le n$, we obtain
\begin{align*}
 p(v)&\le A^{1/(\sigma s+1)}\\
 &\le\frac{1+\sigma/2}{1-2\sigma}\,k(v)\,n^{1/(\sigma s+1)}
 \le\frac{1+2\sigma}{1-2\sigma}\,k(v).
\end{align*}
The last step uses $s\ge2\ln n/\sigma^2$, so
$n^{1/(\sigma s+1)}\le e^{\sigma/2}$, and
$(1+\sigma/2)e^{\sigma/2}\le e^\sigma\le1+2\sigma$.
\end{proof}

\myparagraph{Core-number estimates}
Given balanced pressures, we obtain the core-number estimate $\kh(v)$ for
a non-isolated vertex $v$ by rounding
\[
 \frac{1-2\sigma}{1+2\sigma}\,p(v)
\]
downward, reducing it by at most a $\sigma/64$ fraction. For an isolated
vertex, we set $\kh(v)=0$. Section~\ref{sec:approx} describes how to perform
this rounding.

Theorem~\ref{thm:accuracy} and the
downward rounding of the reported estimate give $\kh(v)\le k(v)$ and, for a
non-isolated vertex,
\[
 \frac{k(v)}{\kh(v)}\le
 \frac{2(1+2\sigma)^2}{(1-2\sigma)(1-\sigma/64)}
 \le2+16\sigma=2+\eps.
\]
The last inequality holds for $0<\sigma\le1/16$. Isolated vertices have
$\kh(v)=0=k(v)$.

\section{Maintaining Balanced Pressures}\label{sec:ds}

We now show how to maintain balanced pressures after each batch, so that
the core-number guarantee of Section~\ref{sec:accuracy} continues to hold.
Our algorithm restores balance through small pressure changes made in
parallel rounds. To prove our main theorem (Theorem~\ref{thm:main}) and obtain a good parallel batch-dynamic algorithm with low depth and work we must bound the number
of rounds per batch and the total work over a sequence of batches.

\myparagraph{Overview} We first define approximate loads and bound their error in
Section~\ref{sec:approx}. We use these error bounds to choose movement and
stopping thresholds. The thresholds ensure that pressure changes have the
correct direction and that stopping guarantees balance for the true loads.
Section~\ref{sec:batch} gives the batch
algorithm, which uses these thresholds to decide which pressures to change and
when to stop. The movement tests use a tighter tolerance than the stopping
test. Section~\ref{sec:rounds} uses this gap and the small size of each
pressure change to bound the number of rounds.

Section~\ref{sec:store} describes the vertex data and the
\textsc{Relabel} operation for maintaining approximate loads as pressures
change. We aggregate neighbor contributions to bound each move's work in
terms of the moving vertex's load. Section~\ref{sec:parallel} gives the
parallel implementation and proves the work and depth bounds.
Section~\ref{sec:work} then relates the repair work to the decrease of a
nonnegative potential. The potential starts at zero, and each batch adds
only $O(1)$ per updated edge, which
bounds how much it can decrease over all repair rounds. This gives the
amortized work bound. Section~\ref{sec:proof} combines these results to
complete the proof of Theorem~\ref{thm:main}.

\subsection{Approximate Loads}\label{sec:approx}

Changing a vertex's pressure can change the load at every neighbor. To
reduce the work of maintaining loads, we round edge shares near one up to one
and sufficiently small shares down to zero. While a share remains in one
of these rounding ranges, its contribution to the approximate load stays
constant, so the algorithm need not update that contribution.

We first describe how integer pressure indices represent the
pressures, and how raises and lowers change them. We then specify the truncation cutoffs and account for numerical
rounding. The resulting error bound for the computed ratio of load to
pressure will justify the balance tests in Section~\ref{sec:batch}.
Section~\ref{sec:store} explains how to maintain the approximate loads
efficiently.

\myparagraph{Pressure indices}
Our algorithm enforces that every pressure is exactly $1/4$
times an integer power of a fixed \emph{step size} $\beta>1$. We choose
$\beta$ in Section~\ref{sec:batch}. This subsection only uses that $2$ is
an integer power of $\beta$. Each vertex $v$ therefore has an integer \emph{pressure index}
$j(v)\ge0$ with $p(v)=\beta^{j(v)}/4$. The pressure index is always an
integer, and it is what our data structure stores. The pressure itself is
determined by the index.

Initially, every vertex has pressure index $0$, so its pressure is $1/4$.
The algorithm changes a pressure only by changing its index. A
\emph{raise} increases $j(v)$ by one, which multiplies the pressure by
$\beta$, and a \emph{lower} decreases $j(v)$ by one, which divides the
pressure by $\beta$. Either change is called a \emph{move}. The batch
algorithm of Section~\ref{sec:batch} also changes pressures in two other
ways, called jumps and resets, and both again produce an integer index
(Section~\ref{sec:store}).

The shares of an edge $uv$ depend only on the difference of the two
pressure indices, because the ratio in
Equation~\eqref{eq:pressure-shares} is
$(p(u)/p(v))^s=\beta^{s(j(u)-j(v))}$.

\myparagraph{Truncating the edge shares}
We compute approximate loads by \emph{truncating} the edge
shares, setting shares near one to one and sufficiently small shares to zero.
For an edge $uv$ with index difference $d=j(u)-j(v)$, use
the share of $v$ except in two cases:
count it as $1$ when $\beta^{sd}>2^{\ceil{\log(64/\sigma)}}$, and as $0$
when $\beta^{sd}<2^{-\ceil{\log(64/\sigma)}-\ceil{\log(4n)}}$. Let $\ellh(v)$ be the sum of these approximate shares.
The two cutoffs serve different purposes. Rounding a share near one causes
only a small relative error. Ignored shares must be smaller by another factor
of $n$, because a vertex can have nearly $n$ of them.

The next lemma shows that the approximate load
can exceed the true load by only a small fraction of the true load, or fall
below it by only a small additive amount independent of the vertex's degree.

\begin{lemma}\label{lem:approx}
Every vertex satisfies
$\ell(v)-\sigma/512\le\ellh(v)\le(1+\sigma/64)\ell(v)$.
\end{lemma}
\begin{proof}
A share rounded to one equals $1-\frac12\beta^{-sd}$, so it
increases by less than
$\frac12 2^{-\ceil{\log(64/\sigma)}}\le\sigma/128$, at
most $\sigma/64$ times that share. An ignored share is smaller by another
factor of $2^{\ceil{\log(4n)}}\ge4n$, so all ignored shares together sum to
less than $\sigma/512$.
\end{proof}

\myparagraph{Rounding}
To compute pressures and shares in finite precision, we store rounded
values of the powers $\beta^i$ with $1\le\beta^i<2$ and scale
them by powers of two as needed. The same table serves both quantities
because $p(v)=\beta^{j(v)}/4$, and every share is determined
by $\beta^{s(j(u)-j(v))}$, which is again a power of $\beta$.

We choose the table's precision so that each computed share differs from
its value after truncation by at most a $\sigma/64$ fraction. The table
approximates each pressure from below, with relative error at most
$\sigma/64$. We use this rounded value in the balance tests below. For a
non-isolated vertex, multiplying it by $(1-2\sigma)/(1+2\sigma)$ gives the
core-number estimate. All sums and balance tests use exact integer
arithmetic, so they introduce no further rounding error
(Appendix~\ref{app:precision}).

\myparagraph{Approximate ratios}
We define the \emph{approximate ratio} used by the algorithm as
\[
 \rh(v)=\frac{\text{computed value of }\ellh(v)}
 {\text{computed value of }p(v)}.
\]

The following lemma bounds the error in this ratio, including both
share truncation and numerical rounding.

\begin{lemma}[Accuracy of the approximate ratios]\label{lem:tests}
At every non-isolated vertex $v$, the approximate ratio satisfies
\begin{equation}\label{eq:test-accuracy}
 \begin{aligned}
 \rh(v)&\le(1+\tfrac\sigma{16})\,r(v),\\
 \rh(v)&\ge(1-\tfrac\sigma{16})\,r(v)
 \qquad\text{if }r(v)\ge\tfrac12.
 \end{aligned}
\end{equation}
\end{lemma}

\begin{proof}
Combining truncation and rounding gives
\[
 \rh\le\frac{(1+\sigma/64)^2}{1-\sigma/64}\,r
 \le(1+\sigma/16)r.
\]
If $r\ge1/2$, then $\ell\ge p/2\ge1/8$, so the truncation error $\sigma/512$ is
at most $(\sigma/64)\ell$. The computed pressure is at most $p$, giving
$\rh\ge(1-\sigma/64)^2r\ge(1-\sigma/16)r$.
\end{proof}

\subsection{Processing a Batch}\label{sec:batch}

We now describe how to process a batch of edge insertions and deletions.
Before inserting edges, the algorithm performs ``jumps'' to quickly raise pressures where needed to
avoid a long sequence of minute increases at vertices receiving many new
edges. 
Once the edge updates are complete, it restores balance through pressure changes in parallel rounds. 
We first bound the effect of a single pressure move and define the balance tests in the algorithm, then give the full batch algorithm.

\myparagraph{Step size}
We set the step size of Section~\ref{sec:approx} to
\[
 \beta=2^{1/(s\ceil{8/\sigma})}.
\]%
By (S3), simultaneous moves by the two endpoints change each share by a
factor of at most $\beta^{2s}$. With our choice,
$\beta^s=2^{1/\ceil{8/\sigma}}\le2^{\sigma/8}$, so this factor is at most
$2^{\sigma/4}$.

We prove in Lemma~\ref{lem:round} that, with these choices, a vertex that
raises its pressure still has load above pressure after the round, and
one that lowers its pressure still has load below pressure. The round and
work analyses use the following bounds:
\begin{equation}\label{eq:params}
 \begin{gathered}
 \beta^{s+1}\le e^{\sigma/8},\qquad
 \beta^{2s+1}\le e^{\sigma/4},\\
 \frac1{\ln\beta}=O(\eps^{-3}\log n).
 \end{gathered}
\end{equation}
The first bound holds because $s\ge3$ and $\ln2<3/4$ give
$\ln\beta^{s+1}=(1+\frac1s)\ln2/\ceil{8/\sigma}
\le\frac43\cdot\frac34\cdot\frac\sigma8$. The second follows from
$2s+1\le2(s+1)$. For the third, $1/\ln\beta=s\ceil{8/\sigma}/\ln2$, and
$\sigma=\eps/16$ gives $s=O(\eps^{-2}\log n)$ and
$\ceil{8/\sigma}=O(\eps^{-1})$.

\myparagraph{Balance tests}
Each round uses the approximate ratio $\rh(v)$ of every
non-isolated vertex $v$ to answer two questions. The first is whether
another round is needed, and the second is which vertices move in it.

We call a non-isolated vertex $v$ \emph{unbalanced} if
$\rh(v)>1+\frac32\sigma$ or $\rh(v)<1-\frac32\sigma$. The algorithm
performs another round whenever some vertex is unbalanced, and stops when
no vertex is.
We choose the tolerance $\frac32\sigma$ because the error in $\rh(v)$ is
less than $\frac12\sigma$ at such a vertex, so every vertex that is not
unbalanced has true ratio within $1\pm2\sigma$, as the balance inequality
of Definition~\ref{def:bfo} requires.

Within a round, more vertices may move than the unbalanced ones. A vertex
that is not unbalanced may still need to move, because the moves of its
neighbors change its load. If only unbalanced vertices moved, a vertex
that is only barely not unbalanced could become unbalanced through its
neighbors' moves, and repairing it could unbalance its neighbors in turn.
Vertices could then keep unbalancing each other back and forth, and we
could not bound the number of rounds. We therefore also move vertices
that are close to being unbalanced.

Precisely, every non-isolated vertex with $\rh(v)>1+\sigma$ raises its
pressure, and every non-isolated vertex with $\rh(v)<1-\sigma$ lowers it.
Section~\ref{sec:rounds} proves two properties that explain these
choices. First, a vertex that does not move in a round cannot become
unbalanced (Lemma~\ref{lem:stay}), so a vertex that is unbalanced after a
round must have moved in that round. This is why we also move vertices
that are not unbalanced, and it leads to the bound on the number of
rounds. Second, although the tests use the approximate ratio $\rh(v)$,
they translate into bounds on the true ratio $r(v)$
(Lemma~\ref{lem:slack}). A vertex raises its pressure only when its true
load exceeds its pressure, and lowers it only when its true load is below
its pressure. A vertex that is not unbalanced has true ratio within
$1\pm2\sigma$.

\myparagraph{The algorithm}
A batch consists of a set $D$ of edges of the graph, which are deleted, and a set
$I$ of edges not in the graph, which are inserted. For a vertex $v$, let
$\deg_I(v)$ be the number of edges of $I$ at $v$.
Lines~\ref{line:batch-jumps-start}--\ref{line:batch-jumps-end} of
Algorithm~\ref{alg:batch} raise pressures where needed to prepare for the
insertions. We call these pressure increases \emph{jumps}
and explain their purpose below. The algorithm then restores balance in
parallel rounds.

We apply the jumps with one call to \textsc{Relabel} and use one
call per repair round to apply all its pressure moves.
This operation
updates the approximate loads along with the pressures.
Section~\ref{sec:store} describes the vertex data and the
\textsc{Relabel} operation.

\begin{algorithm}[htbp]
\caption{Process a batch of edge updates}\label{alg:batch}
\begin{algorithmic}[1]
\Statex \textbf{Input:} Edges $D$ to delete and $I$ to insert.
\State Delete the edges of $D$.\label{line:batch-delete}
\ForAll{vertices $v$ with $\deg_I(v)>0$ \textbf{in parallel}}\label{line:batch-jumps-start}
    \State $p(v)\gets\max\{p(v),2^{\floor{\log\deg_I(v)}}/2\}$\label{line:batch-jump}
\EndFor\label{line:batch-jumps-end}
\State Insert the edges of $I$.
\State Set $p(v)\gets1/4$ for every isolated vertex.\label{line:batch-reset}
\While{some vertex is unbalanced}\label{line:batch-repair}
    \State $O\gets\{v:\deg(v)>0,\ \rh(v)>1+\sigma\}$\label{line:batch-overloaded} \Comment{Vertices to raise}
    \State $U\gets\{v:\deg(v)>0,\ \rh(v)<1-\sigma\}$\label{line:batch-underloaded} \Comment{Vertices to lower}
    \ForAll{$v\in O\cup U$ \textbf{in parallel}}
        \If{$v\in O$}
            \State $p(v)\gets\beta p(v)$ \Comment{\emph{Raise}}
        \Else
            \State $p(v)\gets p(v)/\beta$ \Comment{\emph{Lower}}
        \EndIf
    \EndFor
    \State Update $\ellh$ and $\rh$ at all affected vertices\label{line:batch-update} \Comment{\textsc{Relabel}}
\EndWhile
\end{algorithmic}
\end{algorithm}

Each iteration of the loop at line~\ref{line:batch-repair} is a \emph{round}.
Lines~\ref{line:batch-overloaded}--\ref{line:batch-underloaded} fix the sets
$O$ and $U$ before any pressures change, so all vertices use the ratios at
the start of the round.
Line~\ref{line:batch-update} then updates the approximate loads
and ratios of the vertices that moved and of their neighbors, using the
\textsc{Relabel} operation of Section~\ref{sec:store}. Edge updates and
jumps also update the approximate loads of the affected vertices.
When the repair ends, no vertex is unbalanced. Together with the reset
of isolated vertices on line~\ref{line:batch-reset},
Lemma~\ref{lem:slack}(2) in Section~\ref{sec:rounds}
guarantees balanced pressures, so
Theorem~\ref{thm:accuracy} applies. Some vertices may still have $\rh(v)$ outside $1\pm\sigma$ when repair stops,
but the stopping condition ensures that their loads are close enough to
their pressures for the approximation guarantee to hold.

\myparagraph{Jumps}
A jump on line~\ref{line:batch-jump}
sets $p(v)$ to the larger of its current value and
$2^{\floor{\log\deg_I(v)}}/2$; this target lies between $\deg_I(v)/4$ and
$\deg_I(v)/2$. Jumps are not needed to bound the number of
rounds, which Section~\ref{sec:rounds} bounds without using them. They
keep the total work low. Without jumps, a vertex that receives many new
edges would raise its pressure one step at a time over many rounds, and
each of these rounds would be expensive. For example, suppose we insert an
$n$-vertex clique into an empty graph. Without jumps, its vertices would raise their
pressures from $1/4$ to a constant fraction of the clique size, one round
at a time. Every round would scan all edges of the clique
(Section~\ref{sec:store}), adding a
logarithmic factor to the work bound of Theorem~\ref{thm:main} for fixed
$\eps$. A jump raises $v$'s pressure to a
constant fraction of $\deg_I(v)$ in a single operation.

Deletions need no corresponding step. A vertex that loses many
edges lowers its pressure in the following rounds, and
Section~\ref{sec:work} shows that these lowers are cheap enough.

The following lemma
bounds the pressures and the effect of jumps.

\begin{lemma}\label{lem:facts}
\textup{(1)} Throughout the algorithm, every pressure lies in $[1/4,2n]$.

\textup{(2)} For a batch with insertion set $I$, let $p(v)$ be the pressure
immediately before the jumps and $\bar p(v)$ the pressure immediately after
them. The jumps increase the sum of
pressures by at most $|I|$ and ensure $\deg_I(v)<4\bar p(v)$ for every
vertex $v$. Every vertex that jumps has $p(v)<\deg_I(v)/2$.
\end{lemma}
\begin{proof}
\textup{(1)} All pressures start at $1/4$. Suppose they are all at least
$1/4$ before a round. A vertex that could fall below $1/4$ in a lower has
$p(v)<\beta/4$. Every neighbor then has pressure at least $p(v)/\beta$, so
each incident edge gives $v$ a share of at least $\beta^{-s}/2$.
Since $\beta^{s+1}<2$ by~\eqref{eq:params}, this share exceeds
$\beta/4>p(v)$. Thus a non-isolated vertex at such a pressure has load above pressure,
so it does not lower. Jumps and resets also preserve the lower bound.

For the upper bound, a vertex raises its
pressure only when its load exceeds its pressure. The load is at most the
vertex's degree, which is less than $n$, so the pressure is below $n$
before the raise. The raise multiplies it by $\beta\le2$, so the new
pressure is below $2n$. After a jump, the pressure is at most
$\deg_I(v)/2<n$, so the pressure remains below $2n$.

\textup{(2)} For a vertex with $\deg_I(v)>0$, the jump target satisfies
\[
 \frac{\deg_I(v)}4
 <\frac{2^{\floor{\log\deg_I(v)}}}{2}
 \le\frac{\deg_I(v)}2.
\]
Hence $\bar p(v)>\deg_I(v)/4$, and a vertex
that jumps had $p(v)<\deg_I(v)/2$. The lower bound on $\bar p(v)$ also
holds when $\deg_I(v)=0$, since pressures are positive.
Each vertex's pressure increases by at most $\deg_I(v)/2$. Summing this
bound counts each inserted edge once, giving a total increase of at most
$|I|$.
\end{proof}

Next, we use the bounded range of pressure values to show that each batch requires at
most $O(\eps^{-3}\log^2n)$ repair rounds. Sections~\ref{sec:store}--\ref{sec:parallel}
describe the data structure and bound the cost of these rounds, before
Section~\ref{sec:work} bounds their total work over a sequence of batches.

\subsection{Bounding the Number of Rounds}\label{sec:rounds}

We first prove the two properties of the balance tests that
Section~\ref{sec:batch} announced. We then analyze how one round of
Algorithm~\ref{alg:batch} changes the true ratio of a vertex that moves,
and combine these results to bound the number of rounds.
Lemma~\ref{lem:stay} proves that a vertex that does not move in a round
cannot become unbalanced.

\begin{lemma}\label{lem:stay}
A non-isolated vertex that does not move in a round has true ratio within
$1\pm\frac43\sigma$ after the round, and is not unbalanced.
\end{lemma}
\begin{proof}
Let $v$ be such a vertex, and let $r$ and $r'$ be its true ratios before
and after the round. Since $v$ does not
move, $1-\sigma\le\rh\le1+\sigma$. Since $\rh\ge1-\sigma$,
\eqref{eq:test-accuracy} first gives $r>1/2$. Applying both error bounds
then gives
\[
 \frac{1-\sigma}{1+\sigma/16}\le r
 \le\frac{1+\sigma}{1-\sigma/16},
\]
and hence $|r-1|\le9\sigma/8$. Only the neighbors' pressures change, each
by a factor of at most $\beta$, so (S3) bounds the change in each share of
$v$, and hence in its load, by a factor of
$\beta^s\le e^{\sigma/8}$, using~\eqref{eq:params}.
The pressure of $v$ stays fixed, so its ratio changes by at most the same
factor. Therefore,
\[
 (1-9\sigma/8)e^{-\sigma/8}\le r'
 \le(1+9\sigma/8)e^{\sigma/8}.
\]
These bounds give $|r'-1|<4\sigma/3$. They follow from $e^{-x}\ge1-x$ and
$e^x\le1/(1-x)$ for $0\le x<1$, using $0<\sigma\le1/16$. For example,
\[
 \frac{1+9\sigma/8}{1-\sigma/8}<1+4\sigma/3
 \quad\Longleftrightarrow\quad \sigma(1-2\sigma)>0.
\]
Since $r'\in[1/2,2]$, Lemma~\ref{lem:tests} gives
$|\rh'-r'|\le\sigma/8$ for the approximate ratio $\rh'$ after the round.
Hence $|\rh'-1|<(4/3+1/8)\sigma<\frac32\sigma$, so $v$ is not unbalanced
after the round.
\end{proof}

By the contrapositive of Lemma~\ref{lem:stay}, a vertex that is unbalanced
after a round must have moved in that round. Lemma~\ref{lem:rounds} below
strengthens this statement to show that the vertex moved in the same
direction in every round so far, which bounds the number of rounds.

The tests use the approximate ratio $\rh(v)$, but the rest of the
analysis needs statements about the true ratio $r(v)$.
Lemma~\ref{lem:slack} translates each test into a bound on the true
ratio.

\begin{lemma}\label{lem:slack}
Let $v$ be a non-isolated vertex.

\textup{(1)} If $\rh(v)>1+\sigma$, then $r(v)>1+\frac78\sigma$. If
$\rh(v)<1-\sigma$, then $r(v)<1-\frac78\sigma$.

\textup{(2)} The following implications hold.
\begin{equation}\label{eq:slack}
 \begin{aligned}
 |r(v)-1|\le\tfrac43\sigma&\ \Longrightarrow\ v\text{ is not unbalanced},\\
 v\text{ is not unbalanced}&\ \Longrightarrow\ |r(v)-1|\le2\sigma.
 \end{aligned}
\end{equation}
\end{lemma}
\begin{proof}
We use the error bound~\eqref{eq:test-accuracy} of Lemma~\ref{lem:tests}.
Whenever $1/2\le r\le2$, it gives $|\rh-r|\le\sigma r/16\le\sigma/8$.

\textup{(1)} Suppose $\rh>1+\sigma$. The upper bound
in~\eqref{eq:test-accuracy} gives $r\ge\rh/(1+\sigma/16)>1$. If $r\le2$,
then $r\ge\rh-\sigma/8>1+\frac78\sigma$, and if $r>2$ the claim is
immediate. Now suppose $\rh<1-\sigma$. If $r<1/2$, the claim is immediate,
because $\frac12<1-\frac78\sigma$. Otherwise the lower bound
in~\eqref{eq:test-accuracy} gives $r\le\rh/(1-\sigma/16)<1$, so
$r\le\rh+\sigma/8<1-\frac78\sigma$.

\textup{(2)} For the first implication, $|r-1|\le4\sigma/3$ implies
$r\in[1/2,2]$, so
\[
 |\rh-1|\le|r-1|+|\rh-r|
 \le(4/3+1/8)\sigma<3\sigma/2,
\]
and the vertex is not unbalanced. For the second implication, a vertex
that is not unbalanced satisfies $|\rh-1|\le3\sigma/2$.
Equation~\eqref{eq:test-accuracy} then rules out $r<1/2$ and $r>2$, so the
same error bound applies:
\[
 |r-1|\le|\rh-1|+|\rh-r|
 \le(3/2+1/8)\sigma<2\sigma.
\]
\end{proof}

Part (1) shows that a vertex that raises its pressure has
$\ell(v)>p(v)$ and a vertex that lowers its pressure has $\ell(v)<p(v)$,
so every move brings the pressure toward the true load.
The second implication of~\eqref{eq:slack} shows that, when no vertex is
unbalanced, every non-isolated vertex satisfies the balance inequality of
Definition~\ref{def:bfo}.

We next consider a vertex that moves in a round. We show that a move
brings its true ratio $r=\ell/p$ closer to $1$ without crossing $1$. A
raise therefore leaves load above pressure, and a lower leaves load below
pressure, even when neighboring vertices move in the same round.

\begin{lemma}[One round]\label{lem:round}
Let $p,\ell,r$ denote a vertex's pressure, load, and ratio before a round,
and let $p',\ell',r'$ denote their values after the round.
A vertex that raises its pressure satisfies
$1+\sigma/2<r'\le r/\beta$. A vertex that lowers its pressure satisfies
$\beta r\le r'<1-\sigma/2$.
\end{lemma}
\begin{proof}
When a vertex raises its pressure by a factor of $\beta$, each
neighbor's pressure can increase by at most the same factor. The vertex's
shares therefore cannot increase. Thus $\ell'\le\ell$ and
$r'=\ell'/(\beta p)\le r/\beta$. Similarly, when a vertex lowers its
pressure, its shares cannot decrease, so $\ell'\ge\ell$ and
$r'=\beta\ell'/p\ge\beta r$.

We next show that a raise leaves load above pressure and a lower leaves
load below pressure. A raise requires $\rh>1+\sigma$ and a lower requires
$\rh<1-\sigma$, using the approximate ratio before the round.
Lemma~\ref{lem:slack}(1) therefore gives
$r>1+7\sigma/8$ before a raise and $r<1-7\sigma/8$ before a lower.
By (S3), the load changes by a factor of at most $\beta^{2s}$.
The pressure changes by a factor of $\beta$, so the ratio changes by a
factor of at most $\beta^{2s+1}\le e^{\sigma/4}$, using~\eqref{eq:params}.
Consequently,
\[
 r'>e^{-\sigma/4}(1+7\sigma/8)\ge1+\sigma/2
 \quad\text{for a raise},
\]
\[
 r'<e^{\sigma/4}(1-7\sigma/8)\le1-\sigma/2
 \quad\text{for a lower}.
\]
The last inequality in each line follows from $e^{-x}\ge1-x$
and $e^x\le1/(1-x)$ for $0\le x<1$, using $0<\sigma\le1/16$.
\end{proof}

We next show that a vertex still unbalanced must have raised its
pressure in every round or lowered it in every round. Since each move
changes its pressure by a factor of $\beta$, the pressure range $[1/4,2n]$
limits how many such rounds are possible.

\begin{lemma}[Number of rounds]\label{lem:rounds}
The repair takes $O(\eps^{-3}\log^2n)$ rounds per batch.
\end{lemma}
\begin{proof}
Suppose a non-isolated vertex has $r>1+4\sigma/3$ after $t$ rounds.
Lemma~\ref{lem:round} rules out lowering its pressure in the last round,
and Lemma~\ref{lem:stay} rules out leaving it unchanged. The vertex must therefore have raised its
pressure. Since a raise satisfies $r'\le r/\beta$, its ratio before that
round was even larger. Repeating this argument backward shows that the vertex
raised its pressure in all $t$ rounds.

The same reasoning shows that a vertex with $r<1-4\sigma/3$ after $t$
rounds must have lowered its pressure in every round. A raise would violate Lemma~\ref{lem:round}, an
unchanged pressure would violate Lemma~\ref{lem:stay}, and a lower
satisfies $r'\ge\beta r$, so the ratio before that round was even smaller.

By Lemma~\ref{lem:facts}(1), every pressure stays between $1/4$ and $2n$.
Thus a vertex can change its pressure in the same direction for at most
$\log_\beta(8n)=\ln(8n)/\ln\beta$ consecutive rounds. After more than this
many rounds, every non-isolated vertex must satisfy $|r-1|\le4\sigma/3$.
Equation~\eqref{eq:slack} then guarantees that no vertex is unbalanced,
so the repair must stop.
Substituting $1/\ln\beta=O(\eps^{-3}\log n)$ from
Equation~\eqref{eq:params} gives $O(\eps^{-3}\log^2n)$ rounds.
This bound is deterministic and independent of the batch size.
\end{proof}

\subsection[Vertex Storage and Updates]{Vertex Storage and Updates}\label{sec:store}

We now describe the information stored at each vertex to maintain
approximate loads as pressures change. Our goal is to bound the work of a
move in terms of the moving vertex's load, even when its degree is much
larger. We group neighbors by index, handling larger contributions to the
load individually and combining smaller contributions in counters.

We first express shares and the truncation cutoffs of
Section~\ref{sec:approx} in terms of index differences. Next, we define the neighbor groups and the lists, buckets, and counters
stored at each vertex. We then explain which neighbors a pressure move
visits and how to update the counters when the vertex's index changes.
We finish by combining these updates into \textsc{Relabel},
the operation used for repair rounds and jumps.

\myparagraph{Pressure indices}
Recall from Section~\ref{sec:approx} that each vertex stores its pressure
index $j(v)\ge0$, with $p(v)=\beta^{j(v)}/4$. Let
\[
 K=\ceil{\frac8\sigma},\qquad M=sK,
\]
so that $\beta=2^{1/M}$ and $\beta^s=2^{1/K}$. A raise or lower changes
the index by one. A reset returns it to zero. A jump also sets it to an
integer, because the jump target is
\[
 \frac{2^{\floor{\log\deg_I(v)}}}{2}
 =\frac{\beta^{M(1+\floor{\log\deg_I(v)})}}4.
\]
Lemma~\ref{lem:facts}(1)
gives $0\le j(v)\le\log_\beta(8n)=M\log(8n)$.

\begin{figure}[t]
\centering
\begin{tikzpicture}[x=1.0cm,y=2.1cm]
 \draw[->] (-3.7,0)--(3.9,0);
 \node[below] at (3.9,-0.02) {\small$d$};
 \draw[->] (0,0)--(0,1.18);
 \draw[densely dotted] (-3.7,1)--(3.7,1);
 \draw[densely dotted] (-3.7,0.5)--(3.7,0.5);
 \node[left] at (-3.7,1) {\small$1$};
 \node[left] at (-3.7,0.5) {\small$\frac12$};
 \node[left] at (-3.7,0) {\small$0$};
 \foreach \x/\l in {-3/-3K,-2/-2K,-1/-K,1/K,2/2K,3/3K}
   {\draw (\x,0.02)--(\x,-0.02); \node[below] at (\x,-0.02) {\scriptsize$\l$};}
 \draw[thick,domain=-3.6:0,samples=40] plot (\x,{0.5*exp(0.6931*\x)});
 \draw[thick,domain=0:3.6,samples=40] plot (\x,{1-0.5*exp(-0.6931*\x)});
 \node[above left] at (-0.9,0.3) {\scriptsize $v$ has higher pressure};
 \node[below right] at (0.9,0.7) {\scriptsize $v$ has lower pressure};
\end{tikzpicture}
\caption{The share $f(d)$ of $v$ in an edge $uv$, as a function of the index
difference $d=j(u)-j(v)$.}\label{fig:share}
\end{figure}
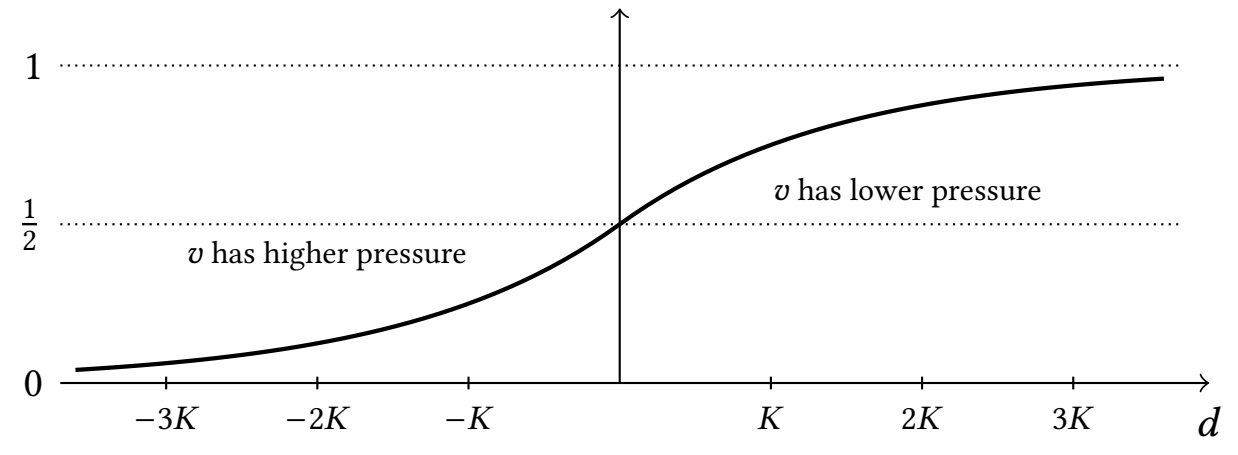

\myparagraph{Shares from indices}
Expressed in terms of the index difference $d=j(u)-j(v)$,
the share of edge $uv$ received by $v$ is
\begin{equation}\label{eq:f}
 x_{uv}(v)=f(d)=\begin{cases}\frac12\,2^{d/K},&d\le0,\\[2pt]
 1-\frac12\,2^{-d/K},&d\ge0.\end{cases}
\end{equation}
This is Equation~\eqref{eq:pressure-shares} with $\beta^s=2^{1/K}$.
Every $K$ additional steps of difference between the indices halves the
edge share of the endpoint with higher pressure (Figure~\ref{fig:share}).
Since $\beta^{sd}=2^{d/K}$, the truncation rule of
Section~\ref{sec:approx} counts a share as $1$ exactly when $d>B$ and as
$0$ exactly when $d<-B-K\ceil{\log(4n)}$, where
\[
 B=K\ceil{\log\frac{64}\sigma}.
\]

\myparagraph{Rounded powers}
In terms of indices, the table of rounded powers from
Section~\ref{sec:approx} holds $2^{i/M}$ for $0\le i<M$. It serves both
pressures and shares because
$p(v)=2^{j(v)/M}/4$ and $2^{d/K}=2^{sd/M}$, using $M=sK$.

\myparagraph{Vertex data}
Fix a vertex $v$ of index $j$. Its \emph{high neighbors} have index at least
$j-B$; the others are \emph{low}. For each high neighbor $u$, vertex $v$
receives a share of at least $f(-B)>\sigma/256$ from edge $uv$, so $v$ has
only $O(\ell(v)/\sigma)$ high neighbors. The next
$K\ceil{\log(4n)}$ index levels below $j-B$ form the \emph{window} of low
neighbors whose contributions are retained.

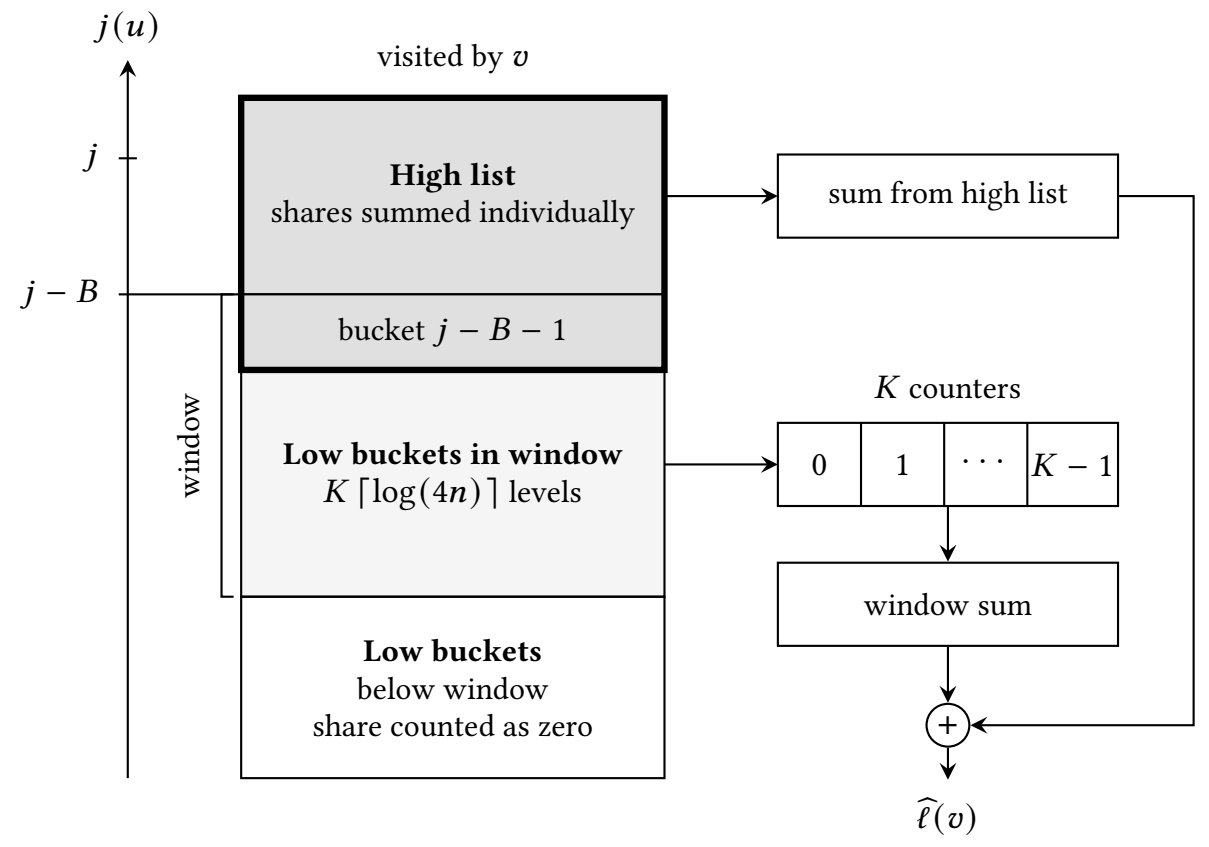
\begin{figure}[htbp]
\centering
\begin{tikzpicture}[x=1cm,y=1cm,font=\scriptsize,>=stealth]
 % Neighbor indices increase upward. The vertical spacing is schematic.
 \draw[->] (0.7,-0.2)--(0.7,4.55);
 \node[above] at (0.7,4.55) {$j(u)$};
 \draw (0.64,3.9)--(0.76,3.9);
 \node[left] at (0.62,3.9) {$j$};
 \draw (0.64,3.0)--(1.45,3.0);
 \node[left] at (0.62,3.0) {$j-B$};

 % All low neighbors have bucket records; only the window has nonzero shares.
 \draw[fill=black!4] (1.45,1.0) rectangle (4.25,3.0);
 \draw[fill=black!12] (1.45,3.0) rectangle (4.25,4.3);
 \node[align=center] at (2.85,3.65)
   {\textbf{High list}\\shares summed individually};
 \draw[fill=black!12] (1.45,2.5) rectangle (4.25,3.0);
 \node at (2.85,2.75) {bucket $j-B-1$};
 \draw (1.4,3.0)--(1.32,3.0)--(1.32,1.0)--(1.4,1.0);
 \node[rotate=90] at (1.1,2.0) {window};
 \node[align=center] at (2.85,1.8)
   {\textbf{Low buckets in window}\\$K\ceil{\log(4n)}$ levels};
 \draw (1.45,-0.2) rectangle (4.25,1.0);
 \node[align=center] at (2.85,0.4)
   {\textbf{Low buckets}\\below window\\share counted as zero};
 \draw[line width=1.2pt] (1.45,2.5) rectangle (4.25,4.3);
 \node[above] at (2.85,4.35) {visited by $v$};

 % The two maintained sums give the approximate load.
 \draw[->] (4.25,3.65)--(5.0,3.65);
 \node[draw,minimum width=2.25cm,minimum height=0.55cm]
   (highsum) at (6.125,3.65) {sum from high list};
 \node[above] at (6.125,2.2) {$K$ counters};
 \draw (5.0,1.6) rectangle (7.25,2.15);
 \foreach \x in {5.55,6.1,6.65}
   {\draw (\x,1.6)--(\x,2.15);}
 \node at (5.275,1.875) {$0$};
 \node at (5.825,1.875) {$1$};
 \node at (6.375,1.875) {$\cdots$};
 \node at (6.95,1.875) {$K-1$};
 \draw[->] (4.25,1.875)--(5.0,1.875);
 \node[draw,minimum width=2.25cm,minimum height=0.55cm]
   (windowsum) at (6.125,0.95) {window sum};
 \draw[->] (6.125,1.6)--(windowsum.north);
 \node[draw,circle,inner sep=1pt] (add) at (6.125,0.15) {$+$};
 \draw[->] (windowsum.south)--(add.north);
 \draw[->] (highsum.east)--(7.75,3.65)--(7.75,0.15)--(add.east);
 \node (load) at (6.125,-0.45) {$\ellh(v)$};
 \draw[->] (add.south)--(load.north);
\end{tikzpicture}
\caption{Storage at $v$, where $j(v)=j$. Low neighbors
outside the window remain in buckets and contribute zero to $\ellh(v)$.
The thick outline marks the high list and first low bucket, scanned by
$v$ on a one-step move.}\label{fig:vertex-storage}
\Description{Neighbor indices are ordered vertically. High neighbors form
a list. Low neighbors occupy index buckets inside and below a retained
window. The window feeds K counters and a window sum, which is added to
the high-list sum to give the approximate load. The high list and top low
bucket are outlined as the neighbors visited by the moving vertex.}
\end{figure}

The vertex stores its index and degree, together with
\begin{itemize}
\item a list of its high neighbors and the sum of their approximate shares;
\item its low neighbors in buckets, one per occupied index, with bucket sizes;
\item $K$ integer counters for the window, described below, and their total
contribution to the approximate load.
\end{itemize}
The two stored sums give $\ellh(v)$, so we can check in constant
time whether $v$ is unbalanced and whether it raises or lowers its pressure.
Each edge has one record at each endpoint; the records point to each other and to their
containing list or bucket.
Figure~\ref{fig:vertex-storage} shows the storage at $v$
and the neighbors it visits when moving one index step.

\myparagraph{Neighbors to visit}
When $v$ changes its index by one, it visits its high list and the bucket just
below it, at index $j-B-1$.
Every other neighbor that does not move stays more than $B$
steps below $v$, so $v$'s approximate contribution at that neighbor remains
one before and after the move.
Its record there remains in the high list and needs no change.
At $v$, these stationary neighbors remain in their buckets,
and the window counters update their aggregate contribution. If a neighbor
$u$ skipped by $v$ also moves, it collects $uv$ through its own high list,
since $v$ has a higher old index. The \textsc{Relabel}
operation below rebuilds the collected edge's records using both
endpoints' new indices.
Notice also that whenever $v$ lies in
a neighbor's low bucket, that neighbor is high at $v$ and is therefore visited.
Thus every bucket record that needs a new index is reached.

\myparagraph{The window counters}
We use the $K$ counters to aggregate the shares received by
$v$ from neighbors in its window. Increasing a neighbor's index by $K$
within the window doubles its contribution to $v$'s load, so we group
levels spaced $K$ apart into the same counter.

We number the window levels from zero at the bottom and
assign a neighbor $u$ at offset $aK+c$, where $0\le c<K$, to counter $c$.
The share received by $v$ from edge $uv$ is
\[
 2^{-B/K-\ceil{\log(4n)}-1}\,2^a2^{c/K}.
\]
The first factor is a power of two common to the whole
window, and $2^{c/K}$ is common to all levels assigned to counter $c$.
Thus counter $c$ stores the sum of the integer weights $2^a$ for the
neighbors at those levels. To recover the window's contribution,
multiply each counter by the table value for $2^{c/K}$, sum, and apply the
common factor. Thus the counters preserve the contributions exactly until
this final table lookup, with no accumulation of rounding error.

A raise shifts the window up one index. Remove the old bottom bucket, whose
members each contribute one to counter $0$. Every remaining offset decreases
by one: move counter $c$ to position $c-1$ for $c>0$, and move counter $0$ to
position $K-1$ and halve it. The halving is exact, because after removing
offset zero all remaining contributions to counter $0$ are even.
A lower reverses the rotation, doubles the counter that wraps around, and adds
the new bottom bucket to counter $0$. Changes at the top of the window are
handled by visiting the high list and the bucket just below it. Recomputing
the window sum takes $O(K)$ work after either shift. For a jump, rebuild the
counters from the bucket sizes in the new window, in $O(K\log n)$ work.

\myparagraph{Relabel}
\textsc{Relabel} takes a set of changing vertices and their
new indices and applies all these changes together. We call it once for
all pressure moves in a repair round or for all initial jumps. It combines
the neighbor visits and counter updates to maintain the stored neighbor
information and approximate loads.
An edge may have two changing endpoints, so we remove its old records
before changing indices and rebuild them using the new indices.

Either every change is by one step, as in a repair round, or all changes
are upward jumps. The following steps run in order, with the work within
each step done in parallel:

\begin{enumerate}
\item[(i)] Each changing vertex scans its high list and the bucket just
below, collecting these edges. Remove duplicates.
\item[(ii)] Remove each collected edge's two records and subtract their
contributions from the stored sums and counters.
\item[(iii)] Change the indices. Shift each changing vertex's counters by one
step, or rebuild them for a jump, using the bucket sizes after (ii).
Recompute its window contribution.
\item[(iv)] Reinsert the collected edges using the new indices, adding their
contributions to the appropriate lists, buckets, counters and sums.
\item[(v)] Retest every vertex whose index or sums changed.
\end{enumerate}
The next lemma shows that the counters account for changes on uncollected
edges, so the stored approximate loads remain correct.

\begin{lemma}\label{lem:correct}
After \textsc{Relabel}, the stored data again represent the current indices
and approximate loads.
\end{lemma}
\begin{proof}
Collected edges are rebuilt explicitly. An uncollected edge cannot have two
changing endpoints: the endpoint with the smaller old index would collect it.
If just one endpoint $v$ changes, the other endpoint starts
at least $B+2$ steps below $v$'s old index. Since $v$ lowers by at most one,
it remains more than $B$ steps above that neighbor.
The record for $v$ stays in that neighbor's high list and
contributes one. At $v$, the neighbor stays in the same low bucket; the counter shift or
rebuild accounts for its new position in the window. Edges with no changing
endpoint need no action.
\end{proof}

\subsection{Parallel Implementation}\label{sec:parallel}

We now give a parallel implementation and bound the work and depth of
edge updates, repair rounds, and jumps. We locate stored records with
dictionaries and combine changes to each sum or counter before updating it.

\myparagraph{Storage}
We store high lists and low buckets in the dictionaries of
Section~\ref{sec:prelims}. Geometric resizing keeps the work of listing
a dictionary proportional to its number of records and adds $O(1)$ expected
amortized work per insertion or deletion. A global table keyed by (vertex, index) locates buckets
and their sizes. Another keyed by edge locates the two endpoint records.
Keep the vertices with $\rh(v)$ outside $1\pm\sigma$ in a table, and the number of
unbalanced vertices in a counter.

\myparagraph{Edge updates}
Insertion adds an edge's two records to their lists or buckets; deletion
removes them. Each record changes one stored sum and possibly one window
counter. Group changes to each sum or counter with a semisort and combine
them with prefix sums, then update degrees and retest the endpoints. A semisort of the
inserted endpoints also gives $\deg_I(v)$. After the insertions, reset isolated
endpoints of deleted edges to index zero; they have no records and their
counters are zero.

\myparagraph{Relabel}
Perform the scans in (i) in parallel and use a semisort to remove duplicates.
Steps (ii) and (iv) are batches of hash-table operations; group and combine
changes to sums and counters as for edge updates. In (iii),
we shift the counters in parallel. For a jump, we rebuild the $K$ counters
in parallel by querying each counter's $O(\log n)$ window buckets
sequentially. This schedule uses $O(K)$ scratch space per changing vertex.
Remove empty buckets after (iv). Step (v) updates the table of vertices with
$\rh(v)$ outside $1\pm\sigma$ and reduces the changes to the counter of unbalanced vertices.
A round uses all vertices in the table and starts only when that counter is
positive.

\begin{lemma}[Cost]\label{lem:cost}
The expected amortized work of \textsc{Relabel} is bounded
by charging
$O((1+\ell(v))/\sigma)$ to each vertex that moves one step, or
$O((\log n+\ell(v))/\sigma)$ to each vertex that jumps. Loads are measured before
the operation. Edge updates cost $O(1)$ expected amortized work
each. Each round takes deterministic $O(\log^2 n)$ depth,
as do the edge updates and jumps of a batch.
\end{lemma}
\begin{proof}
Every scanned neighbor contributes at least
$f(-B-1)\ge f(-B)/2>\sigma/512$, so a changing vertex scans $O(\ell(v)/\sigma)$
edges. Charge each collected edge to a vertex that collected it. Its records
and all resulting tests cost $O(1)$ expected amortized work. Updating counters costs
$O(K)=O(1/\sigma)$ per move, or $O(K\log n)$ per jump. These charges cover all
work, including changes at stationary endpoints.

We group updates and modify records using a constant
number of semisorts, batched dictionary operations, prefix sums and
reductions. Each has an input of polynomial size in $n$ and hence
deterministic $O(\log^2 n)$ depth. Rebuilding a counter after a jump
requires $O(\log n)$ sequential lookups, each with expected constant work
and deterministic $O(\log n)$ depth. Since the $K$ counters
are rebuilt in parallel, the jump also takes $O(\log^2 n)$ depth.
\end{proof}

\subsection{Amortized Work}\label{sec:work}

We now bound the total repair work over a sequence of batches. We relate
the work of each repair round to its decrease in a nonnegative potential that we define shortly.
The potential starts at zero. We show that the deletions, jumps, insertions, and resets
before repair together increase it by only $O(1)$ per updated edge.
This limits the total potential decrease over all repair rounds, which we
use to bound their total work. We defer the proof of the batch increase
bound to Appendix~\ref{app:energy}.

Lemma~\ref{lem:cost} shows that a move of vertex $v$ costs
$O((1+\ell(v))/\sigma)$ expected amortized work. Since $p(v)\ge1/4$, this
cost is $O((p(v)+\ell(v))/\sigma)$. We therefore bound the sum of
$p(v)+\ell(v)$ over all moves, using the values before each round.

\myparagraph{The potential}
For the analysis, allow pressures to be any real numbers at least $1/4$.
Each move adds or subtracts $\ln\beta$ from $\ln p(v)$. We construct a
function whose derivative with respect to $\ln p(v)$ is $p(v)-\ell(v)$.
A raise increases the logarithmic pressure when this derivative is
negative, and a lower decreases it when the derivative is positive.
Define
\[
 F_G(p)=\sum_v p(v)+\sum_{uv\in E(G)}\phi(p(u),p(v)),
\]
\[
 \phi(a,b)=-\ln\min(a,b)
 +\frac1{2s}\left(\frac{\min(a,b)}{\max(a,b)}\right)^s.
\]
This function has the desired derivative,
\begin{equation}\label{eq:gradient}
 \frac{\partial F_G}{\partial\ln p(v)}=p(v)-\ell(v).
\end{equation}
Indeed, for $a\le b$ the derivatives of $\phi$ with respect to $\ln a$ and
$\ln b$ are $-1+\frac12(a/b)^s$ and $-\frac12(a/b)^s$, the negatives of the
two shares. The vertex term therefore contributes $p(v)$, and summing the
edge derivatives gives $-\ell(v)$.

We next show that $F_G$ is strictly convex in logarithmic pressures.
For an edge with endpoint pressures $e^x$ and $e^y$,
\[
 \phi(e^x,e^y)=-\frac{x+y}{2}
 +\frac{|x-y|}{2}+\frac{e^{-s|x-y|}}{2s}.
\]
The first term is linear. The remaining two terms depend only on $t=x-y$,
and their derivative is $\frac12\operatorname{sign}(t)(1-e^{-s|t|})$.
This derivative is nondecreasing, so the edge term is convex. The vertex
terms are exponentials in separate coordinates, so their sum is strictly
convex, and hence so is $F_G$. The vertex terms also dominate the edge terms
as pressures grow, so $F_G$ attains a unique minimum.
Write $p_G^*$ for the pressures at this minimum. Our \emph{potential} is
the nonnegative difference
\[
 F_G(p)-F_G(p_G^*)\ge0.
\]
The algorithm does not compute $p_G^*$. During a repair, the graph and its
minimum are fixed, so the potential decreases by exactly the same amount
as $F_G$.

\begin{lemma}[Potential decrease]\label{lem:drop}
A repair round decreases the potential by
\[
 \Omega\left(\sigma\ln\beta\sum_{v\text{ moves}}(p(v)+\ell(v))\right),
\]
where pressures and loads are taken before the round.
\end{lemma}
\begin{proof}
Use primes for the state after the round. We apply convexity in logarithmic
pressures at $p'$ to obtain a lower bound on $F_G(p)-F_G(p')$.
Substituting~\eqref{eq:gradient} gives
\begin{align*}
 F_G(p)-F_G(p')
 &\ge\sum_v(p'(v)-\ell'(v))\ln\frac{p(v)}{p'(v)}\\
 &=\ln\beta\sum_{v\text{ moves}}|p'(v)-\ell'(v)|.
\end{align*}
For the equality, vertices whose pressures do not change contribute zero.
A raise has $\ln(p/p')=-\ln\beta$ and still has $\ell'>p'$, by
Lemma~\ref{lem:round}. A lower has $\ln(p/p')=\ln\beta$ and still has
$p'>\ell'$. Thus each move contributes $\ln\beta\,|p'-\ell'|$.

We now bound $|p'-\ell'|$ using the pressure and load before the round.
For a vertex that raises its pressure, Lemma~\ref{lem:round} gives
$\ell'-p'>\sigma\ell'/(2+\sigma)$. By (S3) and~\eqref{eq:params},
$\ell'\ge\beta^{-2s}\ell\ge\ell/2$. Since $\ell>p$ before the raise,
we obtain $\ell'-p'=\Omega(\sigma(p+\ell))$.

For a vertex that lowers its pressure, the same lemma gives
$p'-\ell'>\sigma p'/2$. We have $p'=p/\beta\ge p/2$ and $p>\ell$
before the lower, so again $p'-\ell'=\Omega(\sigma(p+\ell))$.
Multiplying these bounds by $\ln\beta$ and summing over all moves
in the round proves the result.
\end{proof}

\myparagraph{Potential increase before repair}
When the graph changes, the minimum value subtracted from $F_G$ also
changes. To bound the increase in potential, we first establish properties
of the minimizing pressures $p_G^*$.

At the minimum, an isolated vertex has pressure $1/4$ because its only
term in $F_G$ is its own pressure. A non-isolated vertex cannot have
pressure $1/4$ at the minimum. Its load would be at least $1/2$, making
the derivative in~\eqref{eq:gradient} negative, so raising its pressure
would decrease $F_G$. Its minimizing pressure is therefore strictly above
$1/4$, so the derivative with respect to its logarithmic pressure must
vanish. Hence its pressure equals its load. Since the loads sum to the
number of edges, we obtain
\begin{equation}\label{eq:mass}
 \sum_v p_G^*(v)=|E(G)|+\tfrac14|\{v:\deg_G(v)=0\}|.
\end{equation}

\begin{lemma}[Pressures at the minimum]\label{lem:minimum}
\textup{(1)} If $p$ is balanced, then every vertex $v$ satisfies
\[
 (1-2\sigma)p(v)\le p^*_G(v)\le(1+2\sigma)p(v).
\]

\textup{(2)} If $G\subseteq G'$ on the same vertex set, then
$p^*_G(v)\le p^*_{G'}(v)$ for every vertex $v$, and
\[
 \sum_v\bigl(p^*_{G'}(v)-p^*_G(v)\bigr)\le |E(G')|-|E(G)|.
\]
\end{lemma}
\begin{proof}
\textup{(1)} For the upper bound, choose a vertex $v$ that maximizes
$p^*_G(v)/p(v)$.
Every neighbor $u$ satisfies
\[
 \frac{p^*_G(u)}{p^*_G(v)}\le\frac{p(u)}{p(v)}.
\]
The share rule therefore gives $v$ no more load under $p_G^*$ than under
$p$. If $v$ has an edge, its pressure at the minimum equals its load, so
\[
 p^*_G(v)\le\ell(v)\le(1+2\sigma)p(v).
\]
If $v$ is isolated, both pressures are $1/4$. Since $v$ maximizes the ratio,
this proves the upper bound at every vertex.

For the lower bound, choose $v$ minimizing $p^*_G(v)/p(v)$. The neighbor
pressure ratios now satisfy the reverse inequality, so $v$ receives at
least as much load under $p_G^*$ as under $p$. For a non-isolated vertex,
this gives
\[
 p^*_G(v)\ge\ell(v)\ge(1-2\sigma)p(v).
\]
An isolated vertex again has pressure $1/4$ under both assignments.
This bounds the minimum ratio and proves the lower bound at every vertex.

\textup{(2)} Suppose some minimizing pressure decreases when edges are
added. Choose $v$ with the smallest ratio $p^*_{G'}(v)/p^*_G(v)$, which
is then less than $1$. The vertex $v$ must have an edge in $G$, since an
isolated vertex already has the smallest allowed pressure, $1/4$.
By the choice of $v$, each old neighbor's pressure relative to $v$'s is at
least as large under $p_{G'}^*$ as under $p_G^*$. Thus no old edge gives
$v$ less load, and new edges can only add load. At both minima, $v$'s
pressure equals its load, so $p^*_{G'}(v)\ge p^*_G(v)$, a contradiction.
The total increase bound follows from~\eqref{eq:mass}, because adding
edges cannot create isolated vertices.
\end{proof}

The next lemma uses these properties to bound the potential increase from
the deletions, jumps, insertions, and resets that precede repair. Its proof
is in Appendix~\ref{app:energy}.

\begin{lemma}[Potential increase from a batch]\label{lem:inject}
For a batch starting from balanced pressures,
lines~\ref{line:batch-delete}--\ref{line:batch-reset} of
Algorithm~\ref{alg:batch} increase the potential by at most
$O(|D|+|I|)$.
\end{lemma}

Starting from balanced pressures lets us bound the potential increase
from deletions by $O(|D|)$. Jumps add at most $|I|$, and resets only
decrease the potential. For insertions, jumps keep each vertex's number
of new incident edges below four times its pressure. This bounds the
potential increase by $O(|I|)$ plus a constant times the total increase
in minimizing pressures, which Lemma~\ref{lem:minimum}(2) bounds by $|I|$.

\begin{corollary}[Work budget]\label{cor:budget}
Over any sequence of $T$ edge updates starting from the empty graph, the
sum of $p(v)+\ell(v)$ over all moves is $O(T/(\sigma\ln\beta))$. The pressure and
load for each move are evaluated before its round.
\end{corollary}
\begin{proof}
Initially the graph is empty and all pressures are $1/4$, so the potential
is zero. Lemma~\ref{lem:inject} bounds the total increase from batches by
$O(T)$. Since the potential never becomes negative, the total decrease
during repairs is also at most $O(T)$. Lemma~\ref{lem:drop} then bounds
the sum of $p(v)+\ell(v)$ over all moves by $O(T/(\sigma\ln\beta))$.
\end{proof}

Combining this corollary with the cost of a move gives
\[
 O\left(\frac{T}{\sigma^2\ln\beta}\right)
 =O(T\eps^{-5}\log n)
\]
expected work for the repair moves, using~\eqref{eq:params}.
Section~\ref{sec:proof} combines this bound with the costs of jumps and edge updates.

\subsection{Proof of Theorem~\ref{thm:main}}\label{sec:proof}

\myparagraph{Accuracy}
The stopping condition and the reset of isolated vertices ensure balance
after every batch. Section~\ref{sec:accuracy} therefore gives
$k(v)/(2+\eps)\le\kh(v)\le k(v)$ for every vertex.

\myparagraph{Work}
The expected amortized cost of a move is
$O((p(v)+\ell(v))/\sigma)$ by Lemma~\ref{lem:cost}, since $p(v)\ge1/4$.
Corollary~\ref{cor:budget} therefore bounds the total
expected amortized work of the moves by
\[
 O\left(\frac{T}{\sigma^2\ln\beta}\right)=O(T\eps^{-5}\log n).
\]

A jumping vertex has $\ell(v)\le(1+2\sigma)p(v)<\deg_I(v)$:
it was balanced or isolated before the batch, deletions only lower loads,
and Lemma~\ref{lem:facts}(2) gives $p(v)<\deg_I(v)/2$.
Thus the expected amortized work of the jumps is
$O(|I|\sigma^{-1}\log n)$ per batch. Each edge update costs $O(1)$
expected amortized work. Every round has a move to pay for its overhead.
Linearity of expectation over the work of the parallel primitives gives the total work bound.

\myparagraph{Depth and space}
Lemma~\ref{lem:rounds} bounds the number of rounds, and Lemma~\ref{lem:cost}
bounds the depth of each round and of the edge updates and jumps. Together
they give depth
\[
 O\left(\frac{\log n}{\ln\beta}\right)\cdot O(\log^2 n)
 =O(\eps^{-3}\log^4n)
\]
per batch deterministically. Each vertex stores $K+O(1)$ numbers,
each edge two records, and every nonempty bucket contains a record. Thus the
space is $O(m+nK)$ plus the table of $M$ rounded powers.
The primitives use linear scratch space, and jump counters
are rebuilt with $O(K)$ scratch space per vertex. Initialization zeros
the indices and counters in $O(nK)$ work. \qed

\section{Experimental Evaluation}\label{sec:exp}

We implement our algorithm as \bfo{}, for ``dynamic balanced pressures'', and compare it with four baselines.
\begin{itemize}
    \item PLDS and PLDSOpt~\cite{LSYDS22}, the parallel batch-dynamic implementations of the level data structure.
    \item Sun, The sequential algorithm of Sun et al.~\cite{SCS20}.
    \item Static, which recomputes all core numbers from scratch.
\end{itemize}
\bfo{} has guaranteed two-sided error bounds between
$3.35$ and $3.79$ on our graphs, below the $4.2$ guaranteed for PLDS. PLDSOpt
uses fewer levels than PLDS and trades a weaker guarantee for speed. With
batches of $10^6$ updates, \bfo{} is $32\times$ faster than PLDS and
$4.1\times$ faster than PLDSOpt in geometric mean on six social and web
graphs, and $9.2\times$ faster than PLDSOpt on three large graphs with up
to $1.8$ billion edges. Recomputing from scratch becomes faster than \bfo{}
only when a batch changes a large part of the graph, about a fifth of the
edges of \emph{livejournal}.

\begin{table}[t]
\caption{Graphs, after removing self-loops, duplicate edges, and isolated
vertices. The column $k_{\max}$ is the largest core number. The last column is the
guaranteed two-sided error bound of \bfo{} (Appendix~\ref{app:fixed-exponent}).}
\label{tab:graphs}
\centering
\small
\begin{tabular}{@{}lrrrr@{}}
\toprule
Graph & Vertices & Edges & $k_{\max}$ & \bfo{} bound\\
\midrule
\emph{dblp} & 317{,}080 & 1{,}049{,}866 & 113 & 3.35\\
\emph{youtube} & 1{,}138{,}499 & 2{,}990{,}443 & 51 & 3.41\\
\emph{wiki} & 1{,}094{,}018 & 2{,}787{,}967 & 124 & 3.41\\
\emph{stackoverflow} & 2{,}584{,}164 & 28{,}183{,}518 & 198 & 3.46\\
\emph{livejournal} & 4{,}846{,}609 & 42{,}851{,}237 & 372 & 3.50\\
\emph{orkut} & 3{,}072{,}441 & 117{,}185{,}083 & 253 & 3.47\\
\midrule
\emph{brain} & 784{,}262 & 267{,}844{,}669 & 1200 & 3.39\\
\emph{twitter} & 41{,}652{,}230 & 1{,}202{,}513{,}046 & 2488 & 3.71\\
\emph{friendster} & 65{,}608{,}366 & 1{,}806{,}067{,}135 & 304 & 3.79\\
\bottomrule
\end{tabular}
\end{table}

\begin{table*}[t]
\caption{Time per batch in milliseconds, and average and maximum error, for
batches of $10^6$ updates on 60 threads. \emph{Static} is the time to
recompute all core numbers from scratch after a batch, without building the
graph. \emph{Sun} is the sequential algorithm of Sun et al., which processes
the updates one at a time on one thread. Its time is $10^6$ times its
average time per update. Bold marks the fastest system and the lowest
average error among PLDS, PLDSOpt, and \bfo{} in each row. All errors are
two-sided, $\max(\kh(v)/k(v),\,k(v)/\kh(v))$, and the guaranteed error
bounds are $4.2$ for PLDS and $3.35$--$3.79$ for \bfo{}
(Table~\ref{tab:graphs}). On the last three graphs, we ran PLDS
only for Ins and did not run Sun. $^*$Average over the first 1705 of 1807 batches
(Section~\ref{sec:exp-large}).}
\label{tab:main}
\centering
\small
\setlength{\tabcolsep}{4pt}
\begin{tabular}{@{}ll|r|rrrr|rrrr|rrrr@{}}
\toprule
& & \multicolumn{5}{c|}{Time per batch (ms)} & \multicolumn{4}{c|}{Average error} & \multicolumn{4}{c}{Maximum error}\\
Graph & Type & Static & PLDS & PLDSOpt & \bfo & Sun & PLDS & PLDSOpt & \bfo & Sun & PLDS & PLDSOpt & \bfo & Sun\\
\midrule
\csname @@input\endcsname tables/main_rows.tex % primitive input: adds no tokens after \bottomrule
\end{tabular}
\end{table*}

\begin{figure*}[t]
\centering
\includegraphics[width=\textwidth]{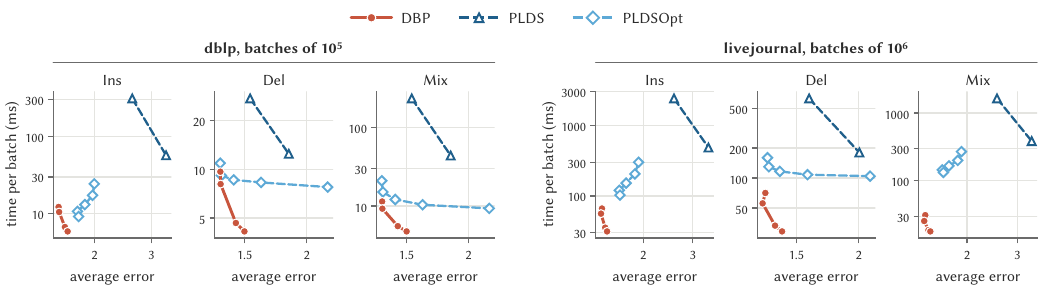}
\caption{Time per batch against average error when varying the parameters,
on \emph{dblp} with batches of $10^5$ and \emph{livejournal} with batches of
$10^6$, on 44 threads. \bfo{} shows four settings with guaranteed error bounds between
$2.6$ and $3.8$. PLDSOpt varies $\delta\in\{0.2,0.4,0.8,1.6,3.2\}$ and PLDS
$\delta\in\{0.4,1.6\}$. Points are connected in the order of their
parameter. Lower and further left is better.}
\label{fig:sweep}
\end{figure*}

\subsection{Implementation}\label{sec:impl}

\bfo{} is written in C++ with ParlayLib~\cite{BAD20}. It diverges from the
theoretical algorithm of Theorem~\ref{thm:main} in its parameters, its repair rule, the
way it stores neighbors, and the estimates it reports. Like the PLDS
implementations~\cite{LSYDS22}, it departs from the analyzed algorithm in some ways in order to obtain speed improvements in practice. We focus on preserving the provable approximation guarantee from our algorithm, however, due to the changes made we do not give a useful bound on the work of \bfo{}.

\myparagraph{Parameters}
Theorem~\ref{thm:main} sets $s=\ceil{2\ln n/\sigma^2}$, which would be extremely large on our graphs. 
Instead, \bfo{} uses $\sigma=0.25$ and the much smaller values $s=28$ and $K=4$ (Section~\ref{sec:store}).
We round a share to one when
the index difference exceeds $bK$ for $b=4$. Appendix~\ref{app:fixed-exponent} proves a maximum error bound for these parameters, which lies between $3.35$ and $3.79$ on our graphs.

\myparagraph{Repair}
Algorithm~\ref{alg:batch} moves every vertex with $\rh(v)$ outside $1\pm\sigma$ by
one step per round. The implementation instead repairs in two monotone
phases. In each round of the \emph{raise phase}, every vertex with $\rh(v)>1+\sigma$
moves to the largest index at which its approximate load is still at least
$1-\tau\sigma$ times its pressure, for $\tau=0.97$. It overshoots its
balance point and ends with $\rh(v)$ just above $1-\sigma$. The phase ends when
no vertex has $\rh(v)>1+\frac32\sigma$. The \emph{lower phase} then moves
vertices with $\rh(v)<1-\sigma$ symmetrically until no vertex has
$\rh(v)<1-\frac32\sigma$. By (S2), raising a vertex cannot decrease the
loads of its neighbors. So in the raise phase $\rh(v)$
drops only when $v$ itself is raised, and each such move leaves
$\rh(v)\ge1-\tau\sigma$. The phase therefore creates no vertex with
$\rh(v)<1-\sigma$. Symmetrically, the lower phase creates no
vertex with $\rh(v)>1+\sigma$. When both phases end, no vertex is unbalanced, so
the accuracy bounds of the balanced pressures framework apply. 
Each phase terminates because indices move in
one direction and are bounded.

\myparagraph{Neighbor storage}
The implementation replaces the structure of Section~\ref{sec:ds}
with a plain array of neighbors per vertex and keeps each vertex's approximate
load as one fixed-point number with 32 fractional bits. 
Thus, a moving vertex
simply scans all its neighbors, recomputes its own approximate load from a table
indexed by index differences, and atomically adds the change of each
neighbor's share to that neighbor's load. A move in our practical implementation therefore costs
$O(\deg(v))$ work instead of $O((1+\ell(v))/\sigma)$. A vertex with at least
$4096$ neighbors avoids most of this cost by listing separately all
neighbors except those far below it, which would contribute nothing to its
approximate load even if it moved down by a fixed margin. The far neighbors
count the vertex with share one, so its moves within the margin do not
change their loads, and a move scans only the list. A neighbor joins the list when it
rises out of the far range, and the list is rebuilt when the vertex moves
down by more than the margin.

\myparagraph{Reported estimates}
The estimates of Theorem~\ref{thm:main} never exceed $k(v)$.
To make the comparison with prior work easier, \bfo{} reports two-sided estimates, and as in~\cite{LSYDS22} the
error of a vertex $v$ with $k(v)>0$ is
\[
 \max\left(\frac{\kh(v)}{k(v)},\,\frac{k(v)}{\kh(v)}\right).
\]
For a non-isolated vertex, \bfo{} reports
$p(v)/\sqrt{c_-c_+}$. Here $c_-$ and $c_+$ depend only on the parameters
and~$n$, and Appendix~\ref{app:fixed-exponent} proves
$c_-k(v)\le p(v)\le c_+k(v)$. So every error is at most
$\sqrt{c_+/c_-}$, the bound listed in Table~\ref{tab:graphs}.

\begin{figure*}[t]
\centering
\includegraphics[width=\textwidth]{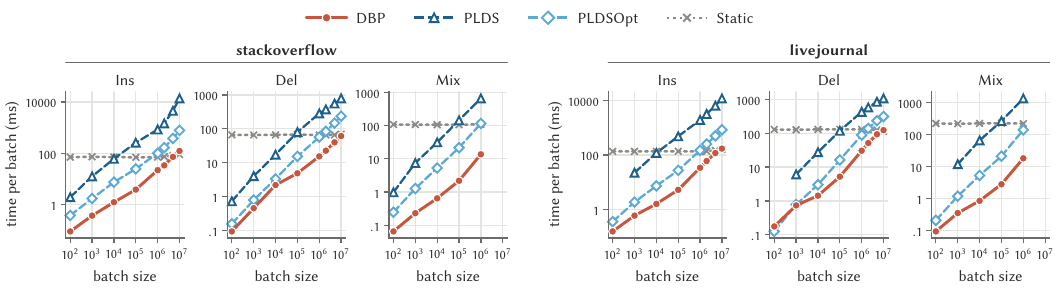}
\caption{Time per batch against batch size on 60 threads. We did not run
PLDS with batches of $10^2$ on \emph{livejournal}.}
\label{fig:batch}
\end{figure*}

\subsection{Setup}

\myparagraph{Machine and measurements}
We use a GCP machine with two 44-core Intel Xeon Platinum 8481C
processors (2.7\,GHz), two-way hyper-threading, and 1.4\,TiB of memory.
Each run uses the cores of one processor, with 60 threads unless stated
otherwise. All code is compiled with \texttt{g++} 15.2 using \texttt{-O3}
and \texttt{-march=native}. The PLDS benchmark driver makes a pass over all vertices after every
batch, and we removed it from the timed region. This makes PLDS and PLDSOpt
up to twice as fast with small batches.

We report the average time per batch over all batches of a run, excluding
the construction of the initial graph. On the six social and web graphs,
the times in Table~\ref{tab:main} and all Static times are medians of three
runs. 
For the larger graphs, we use a single run. 
We compute errors against exact core numbers and report their average and maximum over the vertices with
$k(v)>0$, after the last batch for Ins and Mix and after the middle batch
for Del.

\myparagraph{Graphs and updates}
We use the graphs of~\cite{LSYDS22} in Table~\ref{tab:graphs} and the
update sequences published with PLDS. Six are social and web graphs from
SNAP~\cite{snap}. The three large graphs are \emph{brain}, a dense human
brain network from NeuroData (\url{https://neurodata.io}), \emph{twitter},
the Twitter follower network~\cite{KLPM10} made undirected, and
\emph{friendster}, from SNAP. In \emph{Ins}, all edges are inserted into an
empty graph in batches. In \emph{Del}, all edges are deleted from the full
graph in batches. In \emph{Mix}, one batch with $5\cdot10^5$ insertions and
$5\cdot10^5$ deletions is applied to the graph without the inserted edges.
The updates of \emph{wiki} and \emph{stackoverflow} follow their
timestamps, and those of the other graphs are in random order.

\myparagraph{Algorithms}
All errors in this section are two-sided. In this form, PLDS and Sun et
al.\ guarantee $2+\eps$ with suitable parameters, listed as the one-sided
$4+\eps$ in Table~\ref{tab:comparison}. PLDS uses $\delta=0.4$ and
$\lambda=3$, the setting chosen in~\cite{LSYDS22}, with guaranteed error
bound $(2+3/\lambda)(1+\delta)=4.2$. PLDSOpt uses the same
parameters with $\ceil{\log_{1+\delta}n/50}$ levels per group, which the
PLDS paper describes as trading slightly worse approximation bounds for
speed. For the sequential implementation of Sun et al., we use
$\alpha=1.1$, which gives its most accurate estimates but has no efficiency
guarantee~\cite{LSYDS22}. We omit the exact dynamic algorithms evaluated
in~\cite{LSYDS22}, which were slower than PLDSOpt there on batches of
$10^6$ updates.

\myparagraph{Static recomputation}
We also measure the time to recompute all core numbers from scratch after a
batch, with the parallel exact algorithm of Dhulipala et al.~\cite{DBS17}
and the parallel static approximation of~\cite{LSYDS22} with its default
parameters, both from GBBS. The approximation is $1.1$ to $3.4$ times
faster than the exact algorithm on every input, and we report its time as
\emph{Static}. Static excludes the time to build the graph, which is longer
than the algorithm itself (on \emph{livejournal} Ins, $0.29$\,s against
$0.13$\,s per batch), so it is a lower bound on the cost of recomputing.

\subsection{Batches of a Million Updates}\label{sec:exp-main}

Table~\ref{tab:main} lists the results for batches of $10^6$ updates. On
the six social and web graphs, \bfo{} is the fastest system on every input,
that is, every combination of graph and update type. It is $32\times$ faster than
PLDS in geometric mean (from $5.8\times$ to $100\times$) and $4.1\times$
faster than PLDSOpt (from $1.9\times$ to $10.3\times$). Its maximum errors,
$2.30$--$3.11$, stay within its guaranteed error bounds, while those of PLDSOpt reach
$6$. The same holds after every batch. The largest value over all
batches, $3.34$ on \emph{orkut}, is below its bound of $3.47$. The average error of \bfo{}, $1.09$--$1.65$, is lower than that of
PLDS ($1.45$--$3.33$) on all inputs except \emph{dblp} Del and lower than
that of PLDSOpt ($1.24$--$2.37$) on 14 of the 18 inputs for the six graphs.

The sequential implementation of Sun et al.\ has lower average error than
\bfo{} on every input, $1.00$--$1.21$, but takes $8.7$ to $30\,\mu$s per
update, $420$ to $4300$ times as long as \bfo{}.

Recomputing from scratch competes with \bfo{} only on the three small
graphs, where a batch of $10^6$ updates is between a third and all of the
graph. On \emph{dblp}, \emph{youtube}, and \emph{wiki}, Static takes $0.52$
to $1.45$ times as long as \bfo{}. On \emph{stackoverflow},
\emph{livejournal}, and \emph{orkut}, \bfo{} is $3.2$ to $22$ times faster.
Static is faster than PLDSOpt on 12 of the 18 inputs and faster than PLDS
on all of them. Section~\ref{sec:exp-batch} shows how these comparisons
depend on the batch size.

\begin{table*}[t]
\caption{Vertex moves per update for batches of $10^6$ updates. A move
changes the level of a vertex (PLDS, PLDSOpt) or its index (\bfo{}) by any
number of steps at once.}
\label{tab:moves}
\centering
\small
\setlength{\tabcolsep}{4pt}
\begin{tabular}{@{}l|rrr|rrr|rrr@{}}
\toprule
& \multicolumn{3}{c|}{Ins} & \multicolumn{3}{c|}{Del} & \multicolumn{3}{c}{Mix}\\
Graph & PLDS & PLDSOpt & \bfo & PLDS & PLDSOpt & \bfo & PLDS & PLDSOpt & \bfo\\
\midrule
\csname @@input\endcsname tables/moves_rows.tex % primitive input: adds no tokens after \bottomrule
\end{tabular}
\end{table*}

\myparagraph{Work}
Table~\ref{tab:moves} counts vertex moves per update, a measure of work
that does not depend on how a move is implemented. PLDSOpt moves the fewest
vertices. In geometric mean over the 18 inputs, it makes $0.30$ times as
many moves as \bfo{} and visits $0.27$ times as many neighbors, yet \bfo{}
is $4.1$ times faster. PLDS makes $1.9$ times as many moves as \bfo{} and
$2.3$ times as many neighbor visits but takes $32$ times as long. Most of
the speedup of \bfo{} therefore comes from cheaper moves. A move in PLDS
updates hash tables that store the neighbors of a vertex bucketed by level,
while a move in \bfo{} adds to loads stored in plain arrays
(Section~\ref{sec:impl}).

\subsection{Large Graphs}\label{sec:exp-large}

The last nine rows of Table~\ref{tab:main} are for \emph{brain},
\emph{twitter}, and \emph{friendster}, which have $2.7\cdot10^8$ to
$1.8\cdot10^9$ edges. \bfo{} is faster than PLDSOpt on all nine inputs, by
$9.2\times$ in geometric mean (from $4.6\times$ to $25.8\times$), and $31$
to $94$ times faster than PLDS on Ins. The gaps are largest for Mix, where
PLDSOpt takes $0.27$ to $0.65$\,s for its single batch and \bfo{} $16$ to
$25$\,ms. The average error of \bfo{}, $1.22$--$1.32$, is lower than that
of PLDSOpt ($1.25$--$2.27$) on eight of the nine inputs and lower than that
of PLDS ($3.16$--$3.49$). Its maximum error, $2.55$--$3.02$, stays within
its guaranteed error bounds of $3.39$--$3.79$, while that of PLDSOpt reaches $4.00$.

PLDS needs more memory than the other systems. Its peak resident
memory is $75$\,GiB on \emph{brain} and $1.15$\,TiB on \emph{twitter}, while
\bfo{} uses at most $346$\,GiB and PLDSOpt at most $302$\,GiB. We stopped
PLDS on \emph{friendster} Ins after 1705 of its 1807 batches, when it used
$1.24$\,TiB. Table~\ref{tab:main} reports its average over these batches.

Static takes $0.30$ to $4.9$\,s per batch on these graphs, $7.7\times$ to $208\times$
as long as \bfo{}. PLDSOpt is also faster than Static on all nine
inputs, by $1.3$ to $18$ times, but PLDS is slower than Static on all
three Ins inputs.

\subsection{Accuracy and Running Time}

Figure~\ref{fig:sweep} varies the parameters on \emph{dblp} with batches of
$10^5$ and on \emph{livejournal} with batches of $10^6$, as
in~\cite{LSYDS22}. PLDS and PLDSOpt vary $\delta$ with $\lambda=3$. \bfo{} uses
$\sigma\in\{0.1,0.15,0.25,0.3\}$, with guaranteed error bounds of at most $2.7$,
$2.9$, $3.5$, and $3.8$ on these two graphs. Table~\ref{tab:main} uses
$\sigma=0.25$, and the other settings also change $s$, $K$, and $b$.

The setting of Table~\ref{tab:main} is faster than every setting of PLDS
and PLDSOpt in all six panels, taking $0.14$ to $0.73$ times as long as the
fastest, and it is more accurate than every PLDSOpt setting in three
panels. The setting with the strongest guarantee, $\sigma=0.1$, is more
accurate than every PLDSOpt setting on \emph{livejournal} and takes $0.23$
to $0.68$ times as long as the fastest. On \emph{dblp}, whose batches are
ten times smaller, its average error is at most $0.01$ above that of the
most accurate PLDSOpt setting, and it takes $1.2$ to $1.3$ times as long as
the fastest.

\subsection{Batch Size}\label{sec:exp-batch}

\begin{table}[t]
\caption{Batch size at which recomputing from scratch (Static) becomes
faster than each system, interpolated between the measured batch sizes on a
log scale. A bound marks a system that is faster than Static at the largest
measured batch size, and \emph{never} marks one that is faster even with all
$10^6$ updates of Mix in one batch.}
\label{tab:breakeven}
\centering
\small
\setlength{\tabcolsep}{4pt}
\begin{tabular}{@{}ll|rrr@{}}
\toprule
Graph & Type & PLDS & PLDSOpt & \bfo\\
\midrule
\csname @@input\endcsname tables/breakeven_rows.tex % primitive input: adds no tokens after \bottomrule
\end{tabular}
\end{table}

Figure~\ref{fig:batch} shows the time per batch for batches of $10^2$ to
$10^6$ updates, and up to $10^7$ for Ins and Del. \bfo{} is $7.7$ to $113$
times faster than PLDS at every batch size. It is faster than PLDSOpt at
every batch size on \emph{stackoverflow}, by $1.5$ to $9.9$ times, and on
\emph{livejournal} except Del with batches of $10^2$, where PLDSOpt is
$1.4\times$ faster. On \emph{dblp}, the smallest graph, PLDSOpt is faster
with batches of $10^2$, by up to $3.0\times$, and on Del also with $10^3$.

Static takes about the same time for every batch size, since it processes
the whole graph, while the time of a dynamic system grows with the batch
size. Table~\ref{tab:breakeven} lists the batch size at which each system
becomes slower than Static. PLDS breaks even with batches of
$1.3\cdot10^4$ to $1.2\cdot10^5$ updates, at most $0.3\%$ of the graph, and
PLDSOpt with $5.6\cdot10^5$ to $1.8\cdot10^6$, $2$ to $5\%$. \bfo{} breaks
even only with $5.9\cdot10^6$ to $8.7\cdot10^6$, about a fifth of the
graph. On \emph{stackoverflow} Del, it is still faster than Static with
batches of $10^7$, more than a third of the graph. For Mix, \bfo{} is faster
than Static even with all $10^6$ updates in one batch.

\begin{figure}[t]
\centering
\includegraphics[width=\columnwidth]{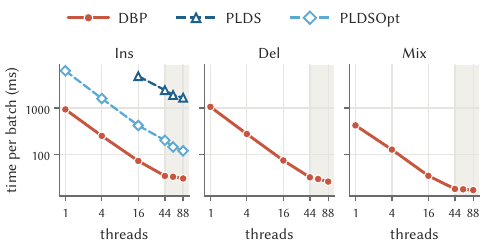}
\caption{Time per batch on \emph{livejournal} against the number of
threads, with batches of $10^6$. We measured PLDS and PLDSOpt only for
Ins. The shaded region uses hyper-threads.}
\label{fig:threads}
\end{figure}

\subsection{Number of Threads}

Figure~\ref{fig:threads} shows the time per batch on \emph{livejournal}
for 1 to 88 threads. With 60 threads, \bfo{} is $27$, $35$, and $23$ times
faster than with one thread for Ins, Del, and Mix. On \emph{dblp}, the
factors are $18$, $20$, and $23$. Beyond 44 threads, the number of cores,
threads share cores through hyper-threading. Going from 44 to 88 threads
reduces the time by $6$--$19\%$ on \emph{livejournal}, and on \emph{dblp}
its effect ranges from $15\%$ slower to $4\%$ faster. PLDSOpt gains more
from parallelism on Ins, a factor of $42$ from 1 to 60 threads, but \bfo{}
is $6.7$ times faster on one thread and faster on 16 threads than PLDSOpt
on 88. On one thread (i.e., measuring empirical work-efficiency), \bfo{} also processes updates $24$ to $44$ times
faster than the sequential implementation of Sun et al.\ on
\emph{livejournal} and $41$ to $109$ times faster on \emph{dblp}.

\section{Conclusion}\label{sec:conclusion}

We gave a batch-dynamic algorithm that maintains a one-sided
$(2+\eps)$-approximation of all core numbers with $O(\eps^{-5}\log n)$
expected amortized work per update and polylogarithmic depth per batch.
To the best of our knowledge, the best previous guarantees were a factor of $4+\eps$ using $O(\log^2 n)$
amortized work.
Under the OMv conjecture no algorithm with
polylogarithmic update and query time can achieve a factor below
two~\cite{CF25}. 

Our algorithm is based around dynamically maintaining balanced pressures.
Edge shares vary smoothly with pressure, so we can restore
balance by making small pressure changes in parallel. We use a potential
argument to bound the total work over any sequence of batches.
The same balanced structure also enables approximating
the densest-subgraph value and supports the applications in
Appendix~\ref{app:applications}.
In our experiments, an implementation with
a smaller guaranteed error bound than PLDS is $32\times$ faster than PLDS and
$4.1\times$ faster than its heuristic variant PLDSOpt in geometric mean on
six social and web graphs, and $9.2\times$ faster than PLDSOpt on three
graphs with up to $1.8$ billion edges.

To keep the presentation simple, we did not try to optimize the dependence of the
work on $\eps$; we know that at least one factor of $1/\eps$ can be removed from the
bound of Theorem~\ref{thm:main} but at the expense of complicating the presentation.
Some interesting questions are to further simplify the framework presented here, and to implement some of the other applications of the general framework explored in the PLDS paper and other orientation papers, which have so far mostly only been studied in theory.
Our practical implementation currently repairs balance with a simple rule that moves
vertices directly past their balance points, to just inside the opposite threshold, in monotone phases. Since it does repair all vertices, it gives provable coreness guarantees, but we have no useful bound on its work.
Such a bound would cover the version of the algorithm that runs fastest in practice.

\bibliographystyle{ACM-Reference-Format}
\bibliography{refs}

\appendix
\section{Omitted Proofs}\label{app:proofs}

\subsection{Additional Details about Primitives}\label{app:primitives}

For small batches, a depth bound that holds with high
probability in a primitive's input size need not hold with high probability
in the graph size $n$. The implementations below give deterministic depth
bounds for every batch size while preserving the expected work bounds. All keys below occupy $O(1)$ words and admit constant-time comparison and
hashing.

\begin{lemma}\label{lem:capped-semisort}
Semisorting $\ell$ items takes $O(\ell)$ expected work, $O(\ell)$ space,
and deterministic $O(\log^2(\ell+2))$ depth.
\end{lemma}
\begin{proof}
We start with the Las Vegas semisort of~\cite{GSSB15}, adapted to binary
forking as in~\cite{BFGS20}. It has linear expected work and space, and
logarithmic depth w.h.p. We run independent attempts on copies of the
input, aborting an attempt before its depth exceeds $C\log(\ell+2)$ or
its space exceeds $C\ell$. For a sufficiently large constant $C$, the
depth bound and Markov's inequality for space ensure that each attempt
succeeds with probability at least $1/2$. We reuse the scratch space between
attempts, clearing it in linear work and logarithmic depth.

After $\ceil{2\log(\ell+2)}$ failed attempts, we finish with deterministic
parallel mergesort. Each merge takes linear work and logarithmic
depth~\cite{BGS10}, so the sort takes $O(\ell\log(\ell+2))$ work,
linear space, and $O(\log^2(\ell+2))$ depth. Each attempt has linear expected work,
the expected number of attempts is at most two, and the probability of
needing mergesort is at most $(\ell+2)^{-2}$. The total expected work is
therefore $O(\ell)$. Both the capped sequence of attempts and the final
sort have deterministic $O(\log^2(\ell+2))$ depth and use linear space.
\end{proof}

\begin{lemma}\label{lem:bounded-dictionary}
A dictionary with $q$ keys uses $O(q+1)$ space. A batch of $\ell$ lookups
takes $O(\ell)$ expected work and deterministic $O(\log(q+\ell+2))$ depth.
A batch of $\ell$ updates takes $O(\ell)$ expected amortized work and
deterministic $O(\log^2(q+\ell+2))$ depth. Updates use $O(q+\ell+1)$
space, and listing all keys takes $O(q+1)$ work and logarithmic depth.
The expected work bounds assume that the stored keys and requested
operations are independent of the dictionary's random choices.
\end{lemma}
\begin{proof}
We hash keys into $C=\Theta(q+1)$ cells using a universal hash function.
Each cell stores a sorted array of record pointers, so records keep their
addresses when the arrays change. Geometric resizing maintains the
capacity bound, with a fresh hash function chosen on each rebuild.
For any fixed lookup key, the expected number of keys in its cell is at
most $1+q/C=O(1)$. Binary search therefore has expected constant work,
deterministic logarithmic depth, and constant scratch space.

For an update batch, we group and combine operations by key using
Lemma~\ref{lem:capped-semisort}, leaving $r\le\ell$ distinct keys.
If $r\le C$, we use the same lemma to group the remaining updates by hash
cell. Within each cell, we sort the updates by key with deterministic
parallel mergesort, merge them with the old array, and filter out deleted
records. If $r_i$ updates enter cell $i$, universality gives
\[
 \mathbb E\!\left[\sum_i r_i^2\right]
 \le r+\frac{r(r-1)}C=O(r).
\]
Hence sorting costs $O(r)$ expected work, since
$r_i\log(r_i+1)=O(r_i^2)$. To bound the work of merging and filtering,
let $A$ be the total number of old records in cells touched by the batch.
A union bound over the updated keys gives
\[
 \mathbb E[A]\le r+\frac{qr}C=O(r).
\]
We merge each touched cell's array with its updates once, so the merges
and subsequent filters take $O(A+r)$ work in total. Since $A\le q$, they
use $O(q+r)$ space. Processing the cells in parallel gives deterministic
$O(\log^2(q+\ell+2))$ depth, even if every key hashes to the same cell.

If $r>C$, we rebuild the whole dictionary by semisorting its records
together with the updates by key, applying the updates, and hashing the
surviving keys into a fresh table of proportional capacity. Here $q=O(r)$,
so semisorting and applying the updates take $O(r)$ expected work. The
collision calculation above gives the same bound for building the sorted
cell arrays. Geometric resizes use the same rebuilding procedure; we
charge their linear expected work to the updates since the preceding
resize. Every rebuild has deterministic $O(\log^2(q+\ell+2))$ depth and
uses linear space. To list the records, we scan the cell directory and its
arrays in $O(q+1)$ work and logarithmic depth.
\end{proof}

In our algorithm, the graph updates and pressure indices determine the
dictionary's keys and requested operations. Random choices affect their
representation and processing order, and exact integer arithmetic ensures
that the order does not change any sum or test. The expected work bounds
therefore apply throughout the update sequence, and the deterministic
depth bounds compose over all rounds.

\subsection{Finite Precision}\label{app:precision}

Use a table of $M$ rational numbers obtained by rounding $2^{i/M}$ down, for
$0\le i<M$, with relative error at most $\sigma/64$. A common denominator that
is a power of two with $O(\log n)$ bits suffices, since $1/\eps\le n^{O(1)}$.
Each entry can be computed with integer arithmetic. Compute $2^{x/M}$ as
$2^{\floor{x/M}}$ times the entry for $x\bmod M$. This gives both pressures and
shares: $p=2^{j/M}/4$, and $2^{d/K}=2^{sd/M}$ because $M=sK$.
Thus pressures round down with relative error at most $\sigma/64$. Multiplying
this value by $(1-2\sigma)/(1+2\sigma)$ gives the reported estimate. A share
$\frac12 2^{d/K}$ has the same relative error. A share $1-\frac12 2^{-d/K}$ for
$d>0$ has absolute error at most $\sigma/128$, hence relative error at most
$\sigma/64$, since it is at least $1/2$.

The window counters use the same table and the power-of-two factor from
Section~\ref{sec:store}. The resulting sum has relative error at most
$\sigma/64$. All computed contributions have
a common power-of-two denominator, so sums can be maintained exactly, without
accumulating rounding error. A window has $K\ceil{\log(4n)}$ levels,
so each neighbor adds $O(n)$ to one counter. Each counter is therefore $O(n^2)$.
The scaling factors need $O(\log n)$ bits as well. Hence every stored number
uses $O(1)$ words and every test is an exact comparison of integers.

The tests and the reported estimates also need $\sigma=\eps/16$ in $O(1)$
words. The tests compare $\rh(v)$ with $1\pm\sigma$ and $1\pm\frac32\sigma$,
and the estimate multiplies the rounded pressure by
$(1-2\sigma)/(1+2\sigma)$. Both are exact because
Theorem~\ref{thm:main} takes $\eps$ to be a rational number whose numerator
and denominator are at most $n^{O(1)}$.

The accuracy guarantee and the bounds of the theorem extend to every
$0<\eps\le1$ with $1/\eps\le n^{O(1)}$. For such an $\eps$, let $\eps'$ be
the largest power of two at most $\eps$. It satisfies the assumption of the
theorem, and $\eps/2<\eps'\le\eps$. Running the data structure with $\eps'$
in place of $\eps$ therefore gives estimates that satisfy the guarantee for
$\eps$, and increases the work, depth, and space bounds by at most constant
factors.

\subsection{Potential Increase from a Batch}\label{app:energy}

We also need a bound on how much an edge term can change. It decreases in each
pressure and satisfies $\phi(ca,cb)=\phi(a,b)-\ln c$. Take
$c=\max(a'/a,b'/b)$, so $a'\le ca$ and $b'\le cb$. Then
$\phi(a',b')\ge\phi(ca,cb)=\phi(a,b)-\ln c$, proving
\begin{equation}\label{eq:edgechange}
 \phi(a,b)-\phi(a',b')\le\ln\max\left(\frac{a'}a,\frac{b'}b\right).
\end{equation}

\begin{proof}[Proof of Lemma~\ref{lem:inject}]
Let $G$, $G_D$, and $G'$ be the graphs before the batch, after deletion, and
after insertion. Let $p$ and $\bar p$ be the pressures before and after the jumps,
and write $\phi_e(p)=\phi(p(u),p(v))$ for the term of edge $e=uv$.

\emph{Deletion.} By Lemma~\ref{lem:minimum}(2), the minimizing pressures decrease
by at most $|D|$ in total. Since edge terms do not increase when pressures rise,
\[
 F_{G_D}(p^*_G)-F_{G_D}(p^*_{G_D})
 \le\sum_v(p^*_G(v)-p^*_{G_D}(v))\le|D|.
\]
Also $p^*_G\ge(1-2\sigma)p$ by Lemma~\ref{lem:minimum}(1),
so~\eqref{eq:edgechange} gives
$\phi_e(p^*_G)-\phi_e(p)\le-\ln(1-2\sigma)$ for each deleted edge. Therefore
\begin{align*}
 \text{potential change}
 &=F_{G_D}(p^*_G)-F_{G_D}(p^*_{G_D})\\
 &\quad+\sum_{e\in D}(\phi_e(p^*_G)-\phi_e(p))\\
 &\le\bigl(1-\ln(1-2\sigma)\bigr)|D|=O(|D|).
\end{align*}

\emph{Jumps.} Pressures increase by at most $|I|$ in total by
Lemma~\ref{lem:facts}(2), and edge terms do not increase. Thus jumps add at
most $|I|$ to the potential.

\emph{Insertion.} Let $\Delta=p^*_{G'}-p^*_{G_D}$. Lemma~\ref{lem:minimum}
gives $\Delta\ge0$, $\sum_v\Delta(v)\le|I|$, and
$p^*_{G_D}\le p^*_G\le(1+2\sigma)p\le(1+2\sigma)\bar p$. Hence
\[
 \ln\frac{p^*_{G'}(w)}{\bar p(w)}
 \le\ln\left(1+2\sigma+\frac{\Delta(w)}{\bar p(w)}\right)
 \le2\sigma+\frac{\Delta(w)}{\bar p(w)}.
\]
Since $F_{G_D}(p^*_{G'})\ge\min F_{G_D}$, insertion adds at most
$\sum_{e\in I}(\phi_e(\bar p)-\phi_e(p^*_{G'}))$ to the potential.
By~\eqref{eq:edgechange}, each summand is bounded by the larger endpoint
log-ratio. Sum the inequality and use $\deg_I(w)<4\bar p(w)$ to get
\[
 \text{potential increase}
 \le2\sigma|I|+\sum_w\frac{\deg_I(w)\Delta(w)}{\bar p(w)}
 \le(4+2\sigma)|I|.
\]

\emph{Isolated vertices.} Resetting their pressures to $1/4$ lowers their only
terms in $F_{G'}$. In total the potential increases by $O(|D|+|I|)$.
\end{proof}

\subsection{Deterministic Sequential Implementation}\label{app:sequential}

The pressure updates and their potential analysis are deterministic. We can
implement them sequentially with the same amortized time bound as the
expected work bound in Theorem~\ref{thm:main}, retaining its accuracy
and space bounds and using the same initialization. Execute the same rounds and the same
remove--change--reinsert phases of \textsc{Relabel}; only the storage and
bookkeeping change.

Replace high lists and the contents of low buckets by doubly linked lists,
using the existing pointers to edge records for constant-time removal.
Keep each vertex's occupied buckets in index order, with a finger search
tree that supports $O(\log n)$ search and $O(1)$ amortized insertion or
deletion at a known position~\cite{HM82}. A record moved by a one-step
round goes to the bucket at its old index or an adjacent index, or to one
of constantly many indices at the high/low boundary. Keep emptied buckets
until the round ends so that old pointers remain valid. A pointer at the
window boundary similarly handles its one-index shifts. Thus ordinary
moves need no general tree searches. Use mark bits to collect each edge
once, update sums directly, and maintain the moving vertices in a linked
list with membership flags. Clear marks only on the records visited.

External edge updates locate records in a balanced search tree keyed by
edge; internal moves use pointers. Updates and jumps may search for
buckets, costing $O(\log n)$ per record. Rebuilding a jumping vertex's
counters visits at most $O(K\log n)$ occupied indices in its new window.
The amortized cost per one-step move is $O((1+\ell(v))/\sigma)$.
An edge update costs $O(\log n)$, and a jump costs
$O((1+\ell(v))\sigma^{-1}\log n)$. As in Section~\ref{sec:proof},
$\ell(v)<\deg_I(v)$ at each jumping vertex, so all jumps cost
$O(|I|\sigma^{-1}\log n)$ per batch. Corollary~\ref{cor:budget} pays for
the ordinary moves exactly as before. The total time for $T$ updates is
therefore $O(T\eps^{-5}\log n)$, deterministically, after initialization.

% Retained in the repository as a separate note, but excluded from the paper.
% \input{sections/worst-case}
\section{Details for the Experiments}\label{app:exp}

\subsection{Accuracy for Fixed Exponents}\label{app:fixed-exponent}

The proof of Theorem~\ref{thm:accuracy} uses the large exponent $s$ only in
its last step. The following version holds for every $s$ and for every
balance tolerance.

\begin{lemma}\label{lem:fixed-exponent}
Let $0<\lambda_-\le1\le\lambda_+$, and suppose that every non-isolated vertex $v$ satisfies $\lambda_-p(v)\le \ell(v)\le\lambda_+p(v)$. Then every such
vertex satisfies
\[
 \frac{k(v)}{2\lambda_+}\ \le\ p(v)\ \le\ \frac{3}{2\lambda_-}\,k(v)\,n^{1/(s+1)}.
\]
\end{lemma}
\begin{proof}
The lower bound is proved as in Theorem~\ref{thm:accuracy}. For the upper
bound, repeat that proof with weight $p^s$ in place of $p^{\sigma s}$. For an
edge with pressures $a\le b$ and $t=(b/a)^s$, the weighted contribution is
\[
 a^s\Bigl(1-\frac1{2t}\Bigr)+b^s\cdot\frac1{2t}
 =a^s\Bigl(1+\frac{t-1}{2t}\Bigr)\le\frac32\,a^s.
\]
The charging argument then gives
\[
 \lambda_-\sum_{u\in S}p(u)^{s+1}\le\frac32\,k\sum_{u\in S}p(u)^s,
\]
and the power mean inequality gives
$p(v)\le\frac{3}{2\lambda_-}k\,|S|^{1/(s+1)}$ as before.
\end{proof}

\myparagraph{The implementation}
The repair of Section~\ref{sec:impl} stops when every non-isolated vertex
has approximate ratio within $1\pm\frac32\sigma$. Its approximate loads
have the cutoffs of Section~\ref{sec:store} with $B=bK$ for a smaller $b$.
A share is counted as one when the index difference exceeds $B$ and is
ignored when the difference is below $-B-K\ceil{\log(4n)}$. Let
$\eta=2^{-b}$. A share counted as one is at least $\frac12$ and is
overcounted by less than $\eta/2$, so by a factor less than $1+\eta$. Each
ignored share is less than $\eta/(8n)$, so less than $\eta/8$ is ignored in
total. Each share is rounded to a multiple of $2^{-32}$, which
changes a load by at most $n2^{-33}$. Since every pressure is at least
$\frac14$, every non-isolated vertex satisfies the hypothesis of
Lemma~\ref{lem:fixed-exponent} with
\[
 \lambda_+=1+\tfrac32\sigma+\tfrac\eta2+4n2^{-33},\qquad
 \lambda_-=\frac{1-\frac32\sigma-4n2^{-33}}{1+\eta}.
\]
The lemma gives the interval $c_-k(v)\le p(v)\le c_+k(v)$ with
$c_-=1/(2\lambda_+)$ and $c_+=3n^{1/(s+1)}/(2\lambda_-)$. Reporting
$p(v)/\sqrt{c_-c_+}$ bounds the two-sided error by $\sqrt{c_+/c_-}$, where $n$
is the number of vertex identifiers of the input. For \bfo{}
($\sigma=0.25$, $s=28$, $b=4$) this bound is between $3.35$ and $3.79$ on the
graphs of Table~\ref{tab:graphs}.

\section{Applications}\label{app:applications}

The applications below use the balanced pressures
maintained by the data structure of Theorem~\ref{thm:main}.

Write $\alpha$ for arboricity (the minimum number of forests
covering the edges) and
$\rho=\max_{\varnothing\ne S\subseteq V}|E(S)|/|S|$ for the maximum subgraph
density; thus $\rho\le\alpha$. Table~\ref{tab:applications} gives bounds for
fixed $\eps>0$ and clique size $k\ge3$. In amortized update bounds,
$\alpha$ denotes the maximum arboricity over the update sequence; forest
and color counts use the current arboricity. Note that none of the applications require knowledge of the
arboricity as input.

\begin{table*}[t]
\caption{Expected amortized work per update for constant accuracy.
Earlier bounds use PLDS~\cite{LSYDS22}, except for density. Forests and
implicit coloring apply the same reductions to PLDS.
Explicit coloring assumes an oblivious adversary. All updates have
polylogarithmic depth; initialization is charged separately.
Our density structure answers value queries in $O(1)$ work
and reports a vertex set $S$ in $O(|S|+\log^2 n)$ work and $O(\log n)$ depth.}
\label{tab:applications}
\centering
\small
\begin{tabular}{@{}lll@{}}
\toprule
Batch-dynamic problem & Our work & Earlier work \\
\midrule
Acyclic orientation, $d^+(v)\le(2+\eps)k(v)$ & $O(\log n)$ & $O(\log^2 n)$, factor $4+\eps$ \\
$(1+\eps)$-approximate densest subgraph & $O(\log n)$ & $O(\log^2 n)$, sequential~\cite{CQ22} \\
Maximal matching & $O(\alpha+\log n)$ & $O(\alpha+\log^2 n)$ \\
Exact $k$-clique count & $O(\alpha^{k-2}\log n)$ & $O(\alpha^{k-2}\log^2 n)$ \\
Partition into $O(\alpha)$ forests & $O(\log n)$ & $O(\log^2 n)$ \\
Implicit $2^{O(\alpha)}$-coloring & $O(\log n)$ & $O(\log^2 n)$ \\
Explicit $O(\alpha\log n)$-coloring & $O(\log n)$ & $O(\log^2 n)$ \\
\bottomrule
\end{tabular}
\end{table*}

\subsection{Densest Subgraph}\label{app:density}

The maximum pressure estimates density more accurately than core numbers.
The reason is that threshold sets replace the peeling order in the weighted
proof of Theorem~\ref{thm:accuracy}.

\begin{lemma}\label{lem:density}
If the graph has an edge and pressures are balanced, then
\[
 \frac{\rho}{1+2\sigma}\le \max_v p(v)
 \le\frac{1+2\sigma}{1-2\sigma}\rho.
\]
\end{lemma}
\begin{proof}
Put $P=\max_v p(v)$. Every edge internal to a densest set contributes one
unit of load to that set. Balance therefore gives
$\rho\le(1+2\sigma)P$.

For the upper bound, set $w(v)=p(v)^{\sigma s}$ at non-isolated vertices
and zero elsewhere. Any nonnegative weights satisfy
\[
 \sum_{uv\in E}\min\{w(u),w(v)\}
 =\int_0^\infty |E(\{v:w(v)\ge t\})|\,dt
 \le\rho\sum_v w(v).
\]
The edge inequality from Theorem~\ref{thm:accuracy} and balance now give,
with both sums over non-isolated vertices,
\[
 (1-2\sigma)\sum_v p(v)^{\sigma s+1}
 \le(1+\sigma/2)\rho\sum_v p(v)^{\sigma s}.
\]
The same power-mean calculation as in that theorem yields
\[
 P\le\frac{1+\sigma/2}{1-2\sigma}\rho\,n^{1/(\sigma s+1)}
 \le\frac{1+2\sigma}{1-2\sigma}\rho.
\]
Here $P$ is attained at a non-isolated vertex, by the lower bound in
Theorem~\ref{thm:accuracy} and the isolated pressure $1/4$.
\end{proof}

Report $(1-2\sigma)/(1+2\sigma)$ times the maximum rounded pressure, or zero
if the graph is edgeless. On a nonempty graph, Lemma~\ref{lem:density}
and rounding give
\[
 \frac{\rho}{1+\eps}\le\widehat\rho\le\rho,
 \qquad
 \frac{\rho}{\widehat\rho}
 \le\frac{(1+2\sigma)^2}{(1-2\sigma)(1-\sigma/64)}
 \le1+8\sigma\le1+\eps.
\]
The penultimate inequality is the same parameter check used for coreness
in Section~\ref{sec:accuracy}, divided by two.

Maintain counts by index and a pointer to the largest occupied index.
A round changes the maximum by at most one; a jump caused by $d$ inserted
edges raises its index by at most $Md$. After isolated vertices reset,
scan down past empty indices. The pointer starts at zero, so upward steps
pay for all downward steps: total extra work is $O(MT)$ plus the number
of moves. Value queries take constant time. The $O(M\log n)$-entry array adds
$O(\eps^{-3}\log^2 n)$ space and initialization work; its scan depth is
absorbed by Theorem~\ref{thm:main}. Update work remains
$O(\eps^{-5}\log n)$ expected amortized, or deterministic sequential time
by Appendix~\ref{app:sequential}.

\myparagraph{Reporting vertices}
On a graph with an edge, put $P=\max_v p(v)$.
For each index $j$, let $S_j=\{v:\deg(v)>0,\ j(v)\ge j\}$, and let
$D=\max_{j:S_j\ne\varnothing}|E(S_j)|/|S_j|$.
Every nonempty set $\{v:w(v)\ge t\}$ with $t>0$ in the proof of
Lemma~\ref{lem:density} is some $S_j$. The upper-bound argument therefore
holds with $D$ in place of $\rho$, giving
\[
 D\ge\frac{1-2\sigma}{1+2\sigma}P
 \ge\widehat\rho\ge\frac{\rho}{1+\eps}.
\]
A reporting query returns the vertices of a set attaining $D$; if the
graph is edgeless, it returns the empty set.

To find this set, we maintain at each index $j$ a dictionary of its
non-isolated vertices, their count $a_j$, and the number $b_j$ of edges
$uv$ with $\min\{j(u),j(v)\}=j$. Then
\[
 |S_j|=\sum_{i\ge j}a_i,\qquad |E(S_j)|=\sum_{i\ge j}b_i.
\]
On a reporting query, suffix sums and a reduction identify the largest
index maximizing the ratio of these counts over nonempty $S_j$. Listing
the vertex dictionaries at or above that index reports the set. There are
$O(M\log n)$ indices, so
the query takes deterministic $O(|S_j|+\eps^{-3}\log^2 n)$ work and
$O(\log n)$ depth. Only reporting queries scan all indices.

Edge insertions and deletions update the edge counts, and index changes
update the vertex dictionaries and counts. Vertices enter or leave these
dictionaries when graph updates change whether they are isolated. During
\textsc{Relabel}, we update the edge counts when collected edges are removed
and reinserted. As in the proof of Lemma~\ref{lem:correct}, an uncollected
edge either has two stationary endpoints or a changing higher endpoint
whose lower endpoint stays fixed. Its minimum endpoint index is therefore
unchanged, so all changes to $b_j$ are covered by edges
already processed. Grouping and combining counter changes with the existing
parallel primitives preserves the update work and depth bounds. The added
dictionaries and counts use $O(n+M\log n)$ space and initialization work.

For fixed $\eps$, this improves the $O(\eps^{-4}\log^2 n)$ amortized
bound of Chekuri--Quanrud~\cite{CQ22}; their logarithmic-time bound allows
an additive error. Pure multiplicative approximations with
polylogarithmic worst-case updates are also known~\cite{CCHHQRSch24,GK25},
and Chekuri et al.~\cite[Lemma 7.1]{CCHHQRSch24} support vertex reporting
in time proportional to the output size. We have not tried to optimize
the additive polylogarithmic overhead of our reporting queries. 

\subsection{Orientation, Matching, and Cliques}\label{app:orientation}

Orient each edge from the smaller pair $(j(v),\mathrm{id}(v))$ to the larger.
This is acyclic, and every outgoing edge contributes at least $1/2$ to its
tail. By balance, Theorem~\ref{thm:accuracy}, and Lemma~\ref{lem:density},
\[
 d^+(v)\le2\ell(v)\le
 \frac{2(1+2\sigma)^2}{1-2\sigma}\min\{k(v),\rho\}
 \le(2+\eps)\min\{k(v),\rho\}.
\]
In particular, maximum outdegree is at most $(2+\eps)\alpha$.

\myparagraph{Reporting changes}
Each reversal appears among the edges collected by \textsc{Relabel}
(Lemma~\ref{lem:correct}). The scan bound and
Corollary~\ref{cor:budget} give $O(T\log n)$ reversals
over $T$ updates. Remember each touched edge's old direction and report its
net change, keeping flips distinct from real edge updates. Remove obsolete
orientation records before adding final ones. Each intermediate graph lies
within the old or new DAG and inherits its outdegree bound. Reporting costs
constant expected work per collected edge.

\myparagraph{Maximal matching}
The PLDS reduction~\cite[Section 9]{LSYDS22} spends constant work per flip
and $O(\alpha)$ work per real update. More precisely, if batch $t$ has size
$b_t$, let $\alpha_t$ be the maximum of one and the arboricities before and
after that batch. The total expected work is
\[
 O\!\left(T\log n+\sum_t b_t\alpha_t\right).
\]
Space is $O(m+n)$. The resulting $O(\alpha+\log n)$ bound matches Blelloch et
al.~\cite{BBDFGH26}; the displayed bound also tracks changes in arboricity.

\myparagraph{Clique counting}
The acyclic-orientation reduction of PLDS~\cite[Section 10]{LSYDS22}
spends $O(\alpha^{k-2})$ work per flip or real update. There are $O(T\log n)$
such events, giving $O(T\alpha^{k-2}\log n)$ expected work and
$O(m\alpha^{k-2}+n)$ space. In particular, triangles take $O(\alpha\log n)$
amortized work. This matches the sequential orientation-based bound
reported in~\cite{DLSY21}. The maximum arboricity is needed for this work
argument: a total flip bound alone does not bound flips weighted by a
changing arboricity.

\subsection{Forests and Colorings}\label{app:coloring}

\myparagraph{Forest partition}
Store each vertex's outgoing edges in compact slots $0,\ldots,d^+(v)-1$.
Edges in slot $i$ form a forest: they have outdegree at most one, so an
undirected cycle would have to be directed, contradicting acyclicity.
A deletion fills its hole with the last edge; batches pair holes with
surviving suffix entries using prefix sums. This costs constant amortized
work per change, giving at
most $\floor{(2+\eps)\alpha}$ forests with $O(\log n)$ expected amortized
update work and $O(m+n)$ space. This is the acyclic case of the
orientation-to-forests reduction of Henzinger et al.~\cite{HNW20}.

\myparagraph{Implicit coloring}
Apply local Cole--Vishkin coloring to each directed forest, as described
in~\cite[Section 1.2]{GG24}. Each color depends on $O(\log^* n)$ successors.
A successor lookup takes constant time. Give sinks color
zero and retain three colors for other vertices. Adjacent vertices differ
in the coordinate of their edge's forest, so the vector of colors is
proper. Coordinates above a vertex's outdegree are zero. This gives
$2^{O(\alpha)}$ colors and $O((1+\alpha)\log^* n)$ query work, with the
forest partition's update and space bounds. Independent queries run in
$O(\log n+\log^* n)$ depth.

\myparagraph{Explicit coloring}
Color each fine level $j$ separately, using disjoint palettes. Put
$p_j=\beta^j/4$. Balance bounds each same-level degree by
$2(1+2\sigma)p_j$, so a fixed degree cap $\ceil{4p_j}$ suffices (using
rounded pressures changes only constants). Apply Ghaffari--Koo's
$(\Delta+1)$-coloring algorithm~\cite[Theorem 1.3]{GK26color}.
For oblivious updates, it uses constant expected work per update and
polylogarithmic depth. Each instance contains all vertices, with those outside
its level isolated. A vertex reports its current level's color.

Every changed edge is an external update or is collected by
\textsc{Relabel}. Deletions precede insertions, preserving each degree cap.
There are $O(T\log n)$ edge and reporting-level changes, determined
independently of the random colors. The coloring theorem therefore gives
$O(T\log n)$ expected work. For largest occupied index $J$, the palette has
$\sum_{j=0}^{J}(\ceil{4p_j}+1)=O(Mp_J)=O(\alpha\log n)$ colors.
The $O(\log^2 n)$ instances store $n$ vertices each, using
$O(m+n\log^2 n)$ space and $O(n\log^2 n)$ initialization work.

\end{document}